\documentclass[12pt]{article}

\usepackage[latin1]{inputenc} 
\usepackage[english]{babel}    
\usepackage[T1]{fontenc}   

\usepackage{amsmath,amssymb} 
\usepackage{amsthm} 

\newtheorem{prop}{Proposition}[section]

\usepackage{listings}
\usepackage{url}
\usepackage{multirow,color}
\usepackage{longtable}
\usepackage{algorithm}
\usepackage{graphicx,caption,subcaption,rotating,setspace,verbatim,algpseudocode,natbib} 
\usepackage{mathtools}
\DeclarePairedDelimiterX{\norm}[1]{\lVert}{\rVert}{#1}

\DeclareMathOperator*{\minimize}{minimize}

\usepackage{psfrag,epsf}
\usepackage{enumerate}

\newcommand{\blind}{0}

\begin{document}

\def\spacingset#1{\renewcommand{\baselinestretch}%
{#1}\small\normalsize} \spacingset{1}


\if0\blind
{
  \title{\bf The Multiple Random Dot Product Graph Model}
  \author{Agnes M Nielsen\\
    Department of Applied Mathematics and Computer Science,\\ Technical University of Denmark\\
    and \\
    Daniela Witten \thanks{
   D.W. was partially supported by NIH Grants  DP5OD009145 and R01GM123993, and NSF CAREER Award
DMS-1252624. }\\
    Departments of Statistics and Biostatistics, University of Washington}
  \maketitle
} \fi

\if1\blind
{
  \bigskip
  \bigskip
  \bigskip
  \begin{center}
    {\LARGE\bf The Multiple Random Dot Product Graph Model}
\end{center}
  \medskip
} \fi

\bigskip
\begin{abstract}
Data in the form of graphs, or networks, arise naturally in a number of contexts; examples include social networks and biological networks. We are often faced with the availability of multiple graphs on a single set of nodes.   In this article, we propose the \emph{multiple random dot product graph model}  for  this setting.  Our proposed model leads naturally to an optimization problem, which we solve using an efficient alternating minimization approach. We further use this model as the basis for a new test for the hypothesis that the graphs come from a single distribution, versus the alternative that they are drawn from different distributions. We evaluate the performance of both the fitting algorithm and the hypothesis test in several simulation settings, and demonstrate empirical improvement over existing approaches. We apply these new approaches to  a Wikipedia data set and a \emph{C. elegans} data set. 
\end{abstract}

\noindent%
\vfill

\newpage
\spacingset{1} 


\section{Introduction}

A graph, or a network,  consists of a set of vertices, or nodes, and the edges between them. Data in the form of graphs arise in many areas of science. Examples include social networks, communication structures, and biological networks \citep{pansiot1998routes,dawson2008study,miller2010divergence}.

 Let $A \in \{0,1\}^{n \times n}$ denote the $n \times n$ adjacency matrix corresponding to an unweighted and undirected graph with $n$ nodes; $A_{ij}=1$ if there is an edge between the $i$th and $j$th nodes, and $A_{ij}=0$ otherwise. A number of models for graphs have been proposed in the literature.
The simplest is the Erd\"os-Renyi model \citep{erdos1959random}, in which all edges are assumed independent and  equally likely: that is, 
$$P(A_{ij}=1)=\pi,$$ for $\pi \in [0,1]$.  The stochastic block model \citep{holland1983stochastic} generalizes the Erd\"os-Renyi model by assuming that each node belongs to some latent class, and furthermore that the probability of an edge between a pair of nodes depends on their latent class memberships. 
Letting $\tau \in \{1,\ldots,M\}^n$ denote the latent class membership vector, and $B \in [0,1]^{M \times M}$ the block connectivity probability matrix, this leads to the model 
\begin{equation}
\label{eq:sbm}
P\left(A| \tau, B\right) = \prod_{i<j} B_{\tau_i, \tau_j}^{A_{ij}} \left(1-B_{\tau_i, \tau_j} \right)^{\left(1-A_{ij} \right)}.
\end{equation}
The stochastic block model is further generalized by  latent space models \citep{hoff2002latent,hoff2008modeling,hoff2009multiplicative,ma2017exploration,wu2017generalized}, which assign each vertex a position in a latent space, and posit that the probability of an edge between two vertices depends on their positions. In particular, in the random dot product graph model \citep{young2007random,nickel2008random,scheinerman2010modeling}, each node is assigned a vector, and the probability of an edge between two nodes is a function of the dot product between the corresponding vectors: that is, 
\begin{equation}
\label{eq:rdpg}
P\left(A_{ij}=1\right) = f\left(x_i^T x_j\right),
\end{equation}
where $x_1,\ldots,x_n \in \mathbb{R}^d$ are $d$-dimensional  latent vectors associated with the $n$ nodes, and $f(\cdot)$ is a link function.  If $f(\cdot)$ is the identity, then this model can be fit via an eigendecomposition.
 The Erd\" os-Renyi model is a special case of the random dot product graph, as is the stochastic block model, provided that the matrix $B$ in \eqref{eq:sbm} is positive semi-definite.
 
The models described above are intended for data that consist of either a single graph, or else multiple graphs that are assumed to be independent and identically distributed draws from a single distribution. 
However, in many contemporary data settings, researchers collect multiple graphs on a single set of nodes \citep{ponomarev2012gene,stopczynski2014measuring,szklarczyk2014string,nelson2017comparison}. These graphs may be quite different, and likely are not independent and identically distributed.   
For instance, if the nodes represent people, then the edges in the two graphs could represent Facebook friendships and Twitter followers, respectively. 
If the nodes represent proteins, then the edges in the two graphs could represent binary interactions (i.e. a physical contact  between  a pair of proteins) and co-complex interactions (i.e. whether a pair of proteins are part of the same protein complex), respectively \citep{yu2008high}. Alternatively, if the nodes represent brain regions, then each graph could represent the connectivity among the brain regions for a particular experimental subject \citep{hermundstad2013structural}. 

Several models have been proposed for this multiple-graph setting \citep{tang2009clustering,shiga2012variational,dong2014clustering,durante2017nonparametric,wang2017common}.  The multiple random eigen graphs (MREG) model \citep{wang2017joint} extends the random dot product graph model: letting $A^1,\ldots,A^K  \in \{0,1\}^{n \times n}$ denote $K$ adjacency matrices on a single set of $n$ nodes, this model takes the form
\begin{equation}
\label{eq:mreg}
P\left(A^k_{ij}=1\right) = f\left(W_{ij}^k\right), \quad W^k = U \Lambda^k U^T, \quad k=1,\ldots,K,
\end{equation} 
where $U$ is a $n \times d$ matrix of which the $i$th row, $u_i = (u_{i1},\ldots,u_{id})^T \in \mathbb{R}^d$, is the $d$-dimensional latent vector associated with the $i$th node across all $K$ graphs; $\Lambda^1,\ldots,\Lambda^K$ are $d \times d$ diagonal matrices; and $f(\cdot): \mathbb{R} \rightarrow [0,1]$ is a link function. If $\Lambda^1=\ldots=\Lambda^K$ and further $W^1=\ldots=W^K$ are positive semi-definite, then \eqref{eq:mreg} reduces to the random dot product graph model \eqref{eq:rdpg}. However, in general, \cite{wang2017joint} does not guarantee that $W^1,\ldots,W^K$ be positive definite. In particular, the model \eqref{eq:mreg} is fit using an iterative approach that estimates one column of $U$ at a time; this algorithm does not guarantee that the columns of $U$ be mutually orthogonal, or that the diagonal elements of $\Lambda^1,\ldots,\Lambda^K$ be nonnegative \citep{wang2017joint}. 

In this paper, we will consider this multiple graph setting. Our contributions are as follows:
\begin{enumerate}
\item We present the \emph{multiple random dot product graph} (multi-RDPG), a refinement of the MREG model \eqref{eq:mreg} of \cite{wang2017joint}. 
This model  provides a more natural generalization of the random dot product graph \eqref{eq:rdpg} to the setting of multiple random graphs, by requiring that the matrices $W^1,\ldots,W^K$ be positive semidefinite. In particular, unlike the proposal of \cite{wang2017joint}, the multi-RDPG with $K=1$ reduces to the random dot product graph model. 
\item We derive a new algorithm for fitting the multi-RDPG model, which   simultaneously estimates all  $d$ latent dimensions, while also enforcing their orthogonality. It therefore  yields improved empirical results relative to the proposal of \cite{wang2017joint}. 
\item We develop a new approach for testing the hypothesis that multiple graphs are drawn from the same distribution. This approach follows directly from the multi-RDPG model. 
\end{enumerate}

The rest of this paper is organized as follows. In Section~\ref{sec:model}, we present the multi-RDPG model, as well as an algorithm for fitting this model. We develop a test for the null hypothesis that two graphs are drawn from the same RDPG model in Section~\ref{sec:test}. Results on a Wikipedia data set and a \emph{C. elegans} connectome data set are presented in Sections~\ref{sec:realdata}. We close with the Discussion in Section~\ref{sec:disc}. 

\section{The Multiple Random Dot Product Graph Model} 
\label{sec:model}

In this section, we will extend the random dot product graph model of \cite{young2007random} to the setting of multiple unweighted undirected graphs, which we assume to be drawn from related though not necessarily identical distributions. 

\subsection{The Multiple Random Dot Product Graph Model}
\label{sec:multi-rdpg}

To begin, we notice from \eqref{eq:rdpg} that in the case of a single unweighted and undirected graph with $n$ nodes, the random dot product graph model  assumes that $P\left( A_{ij}=1 \right) =f\left( W_{ij} \right)$, for a rank-$d$ positive semi-definite $n \times n$ matrix $W \equiv U \Lambda U^T$. Here $U$ is an $n \times d$ orthogonal matrix, such that $U^T U=I$, and $\Lambda$ is a $d \times d$ diagonal matrix with positive elements on the diagonal. The fact that the matrix $W$ is positive semi-definite allows us to interpret the rows of  the $n \times d$ matrix $W^{1/2} = U \Lambda^{1/2}$ as the positions of the $n$ nodes in a $d$-dimensional space, and thus the probability of an edge between a pair of nodes as a function of the distance between the nodes in this $d$-dimensional space. 

To extend this model to the case of $K$ unweighted and undirected graphs, we propose the \emph{multiple random dot product graph} (multi-RDPG) model, which is of the form
\begin{equation}
 P\left( A_{ij}^k=1 \right) =f\left( W_{ij}^k \right), \quad \quad
 W^k = U \Lambda^k U^T, \quad \quad k=1,\ldots,K, \label{eq:mrdpg}
\end{equation}
where $U$ is an $n \times d$ orthogonal matrix, $\Lambda^1,\dots,\Lambda^K$ are $d \times d$ diagonal matrices with nonnegative diagonal elements, and $f(\cdot): \mathbb{R} \rightarrow [0,1]$ is a link function. 

While the multi-RDPG \eqref{eq:mrdpg} appears at first glance quite similar to the MREG formulation \eqref{eq:mreg} of \cite{wang2017joint}, there are some key differences. In particular, the multi-RDPG model constrains the columns of $U$ to be orthogonal, and the diagonal elements of $\Lambda^1,\ldots,\Lambda^K$ to be nonnegative; by contrast, in \eqref{eq:mreg}, these are no such constraints. In greater detail, the differences between the two models are as follows:
\begin{enumerate}
\item For $k=1,\ldots,K$, $W^k$ in \eqref{eq:mrdpg} is a positive semi-definite matrix of rank $d$. Thus, the nodes in the $k$th graph can be viewed as lying in a $d$-dimensional space, such that the probability of an edge between a pair of nodes is a function of their distance in this space.  By contrast, such an interpretation is not possible in the MREG formulation \eqref{eq:mreg}, in which there are no guarantees that the matrix $W^k$ is positive semi-definite. 
\item  With $K=1$, the multi-RDPG model \eqref{eq:mrdpg} simplifies to the random dot product graph model \eqref{eq:rdpg}. The same is not typically true of the MREG model \eqref{eq:mreg}, since the matrix $W^1$ in \eqref{eq:mreg} need not be positive semi-definite. 
\item The multi-RDPG model can be efficiently fitted via a single optimization problem, as detailed in Section~\ref{sec:algo}.
 By contrast, the MREG model \eqref{eq:mreg} is fitted by estimating one dimension at a time \citep{wang2017joint}, which can lead to poor results in estimating the entire subspace $U$ in \eqref{eq:mrdpg}.
\end{enumerate}

\subsection{Optimization Problem}
\label{sec:algo}





To derive an optimization problem for fitting the multi-RDPG model \eqref{eq:mrdpg}, we consider the simplest case, in which $f(\cdot)$ is the identity. To motivate our optimization problem, we consider the case of $K=1$, in which the multi-RDPG model coincides with the random dot product graph model \eqref{eq:rdpg}. The model \eqref{eq:rdpg} is typically fit by solving the optimization problem \citep{scheinerman2010modeling}
\begin{equation}
\minimize_{X \in \mathbb{R}^{n \times d}} \| A - XX^T \|_F^2,
\label{eq:fit-rdpg}
\end{equation}
or a slight modification of \eqref{eq:fit-rdpg} if no self-loops are allowed. Let $A=VDV^T$ denote the eigen decomposition of $A$, where the diagonal elements of $D$ are ordered from largest to smallest. Then, the solution to \eqref{eq:fit-rdpg} is $\hat{X}=V_{[1:d]} \left( D_{[1:d]}^+ \right)^{1/2}$, where $V_{[1:d]}$ is the $n \times d$ matrix that consists of the first $d$ eigenvectors of $A$, and where $D_{[1:d]}^+$ is the $d \times d$ diagonal matrix whose diagonal elements are the positive parts of the first $d$ eigenvalues of $A$. 
Equivalently, we can fit \eqref{eq:rdpg} by solving the problem
\begin{equation}
\minimize_{U \in \mathbb{R}^{n \times d}, \; U^T U=I, \; \Lambda \in \Delta_+^d} \| A_+ - U \Lambda U^T \|_F^2,
\label{eq:fit-rdpg2}
\end{equation}
where $A_+ \equiv VD_+ V^T$, $D_+ \equiv D_{[1:n]}^+$, and $\Delta_+^d$ is the set of diagonal $d \times d$ matrices with nonnegative diagonal elements. It is not hard to show that   $\hat{X}=\hat U \hat\Lambda^{1/2}$.

To fit the multi-RDPG model \eqref{eq:mrdpg}, we directly extend the optimization problem \eqref{eq:fit-rdpg2} to accommodate $K$ adjacency matrices,
\begin{equation}
\minimize_{U \in \mathbb{R}^{n \times d}, \; U^T U=I, \; \Lambda^1,\ldots,\Lambda^K \in \Delta_+^d} \sum_{k=1}^K \| A_+^k - U \Lambda^k U^T \|_F^2.
\label{eq:fit-mrdpg}
\end{equation}

\subsection{ Algorithm}\label{sec:alg}

While the optimization problem \eqref{eq:fit-rdpg2} for fitting the random dot product graph model \eqref{eq:rdpg} has a closed-form solution, its extension to the multi-RDPG model \eqref{eq:fit-mrdpg} does not. 
 Thus, to solve \eqref{eq:fit-mrdpg}, we take an alternating minimization approach \citep{csiszar1984information}, which will rely on the following three results.

 First, we derive a majorizing function \citep{hunter2004tutorial} that will prove useful in solving \eqref{eq:fit-mrdpg}. 
 \begin{prop}
 \label{prop:majorize}
The function
$$ g(U)\equiv -2 \sum_{k=1}^K \mathrm{trace} \left(U \Lambda^k U^T A_+^k\right)$$
 is majorized by
 $$h(U \mid U') \equiv -g(U') -4 \sum_{k=1}^K \mathrm{trace} \left( \Lambda^k  U'^T A_+^kU \right), $$
in the sense that $g(U) \leq h(U \mid U')$ and $g(U')=h(U' \mid U')$. 
\end{prop}

Next, we use Proposition~\ref{prop:majorize} to devise a simple iterative strategy that is guaranteed to decrease the value of the objective of \eqref{eq:fit-mrdpg} each time $U$ is updated. 
\begin{prop}
Let $U^{\mathrm{old}}$ denote an orthogonal $n \times p$ matrix,  and let $B$ and $C$ be the matrices whose columns are the left and right singular vectors, respectively, of the matrix 
$\sum_{k=1}^K   A_+^k U^\mathrm{old} \Lambda^k.$
Then, for 
${U}^{\mathrm{new}} \equiv B C^T$, 
$$ \sum_{k=1}^K \| A^k_+ -  U^\mathrm{new}  \Lambda^k (U^{\mathrm{new}})^T  \|_F^2 \leq  \sum_{k=1}^K \| A^k_+ - U^\mathrm{old} \Lambda^k (U^{\mathrm{old}})^T \|_F^2.$$
\label{prop:updateU}
 \end{prop}

Finally, we show that with $U$ held fixed, \eqref{eq:fit-mrdpg} can be solved with respect to $\Lambda^1,\ldots,\Lambda^K$ for the global optimum. 
  \begin{prop}
If $U^T U = I$, then the solution to 
 \begin{equation}
 \minimize_{\Lambda^1,\ldots,\Lambda^K \in \Delta_+^d} \sum_{k=1}^K \| A_+^k - U \Lambda^k U^T \|_F^2
\label{eq:optLambda}
\end{equation}
is
\begin{equation}
\Lambda^k_{jj} = \max\left(0, Z^k_{jj}\right),
\label{eq:updateLambda}
\end{equation}
where 
$Z^k = U^T A_+^k U$.
\label{prop:updateLambda}
 \end{prop}

 Propositions~\ref{prop:majorize}--\ref{prop:updateLambda} immediately suggest an alternating minimization algorithm for solving \eqref{eq:fit-mrdpg}, which is summarized in Algorithm~\ref{alg:multiRDPG}.

\begin{algorithm}
    \caption{Alternating Minimization Algorithm for Solving \eqref{eq:fit-mrdpg}}
\label{alg:multiRDPG}
    \begin{algorithmic}[1] 
            \State For $k=1,\ldots,K$, initialize $\Lambda^k$ to be a $d \times d$ diagonal matrix with nonnegative diagonal elements.
            \State  Initialize $U^{\text{old}}$ to be an orthogonal $n \times d$ matrix.
            \State  For $k=1,\ldots,K$, let $VDV^T$ denote the eigendecomposition of $A^k$, and define $A_+^k \equiv V D_+ V^T$, where $D_+$ is the diagonal matrix with diagonal elements $(D_+)_{ii} = \max(D_{ii},0)$ for $i=1,\ldots,n$. 
            \While{not converged} 
               \State \parbox[t]{\linewidth}{Define the matrices $B$ and $C$ to have as their columns the left and right singular vectors, respectively, of the matrix $  \sum_{k=1}^K   A_+^k U^\mathrm{old} \Lambda^k$. Then, update $U \leftarrow B C^T$.}\label{alg:multiRDPG_U}
                \State \parbox[t]{\linewidth}{For $k=1,\ldots,K$ and $j=1,\ldots,n$, update $\Lambda^k_{jj}\leftarrow \max(0,Z_{jj})$, where $Z_{jj}$ is the $j$th diagonal element of the matrix $Z=U^T A_+^k U$.}\label{alg:multiRDPG_L}
                \State Update $U^{\text{old}} \leftarrow U$. 
            \EndWhile
    \end{algorithmic}
\end{algorithm}

\subsection{Simulation Study}\label{sec:sim1}
We conducted two simulation studies in order to evaluate the performance of Algorithm \ref{alg:multiRDPG} for fitting the multi-RDPG model \eqref{eq:mrdpg} (multi-RDPG). We compare it to the algorithm of \cite{wang2017joint} for fitting the multiple random eigen graph model \eqref{eq:mreg} (MREG) using software obtained from the authors, the random dot product graph \citep{young2007random} fitted to the average of all of the adjacency matrices (RDPG$_{\text{all}}$), and the random dot product graph \citep{young2007random} fitted to the replicates from each model separately (RDPG$_{\text{separate}}$). The two versions of RDPG are fitted using \verb=R= base functions \citep{team2015r}.

 To quantify the error in estimating the matrix $U$, we made use of subspace distance  \citep{absil2006largest}, defined as
\begin{equation}
d_U(\hat{U},U) \equiv ||P_{\hat{U}}-P_{U}||_2, \label{eq:subspacedist}
\end{equation}
where $P_A \equiv A(A^TA)^{-1}A^T$, the notation $\| \cdot \|_2$ indicates the matrix 2-norm, and $\hat{U}$ is an estimate of the matrix $U$.
To quantify the error in estimating $\Lambda^1,\ldots,\Lambda^K$, we computed the adjacency matrix error,
\begin{equation}
d_{A}(\hat{\Lambda},\Lambda) \equiv \frac{1}{K} \sum_{k=1}^K \left\|f(U \Lambda^k U^T) - \hat{U} \hat{\Lambda}^k \hat{U}^T\right\|_F^2, \label{eq:lambdadist}
\end{equation}
 where $f(\cdot)$ is the link function defined in \eqref{eq:mrdpg}, and where the notation $\| \cdot \|_F$ indicates the Frobenius norm. 

\subsubsection{Simulation Setting 1}
The first simulation setting is based upon the simulation study in \cite{wang2017joint}. 
We set $n=20$ and $d=3$.  We defined $U\in \mathbb{R}^{n\times d}$ to be the orthogonal matrix with columns
\begin{align}
U_{\cdot 1} & = \begin{pmatrix} 1 & 1 & 1 & 1&  \ldots & 1 \end{pmatrix}^T/\sqrt{n}\nonumber \\
U_{\cdot 2} & = \begin{pmatrix} 1 & -1 & 1 & -1 & \ldots & 1 & -1 \end{pmatrix}^T/\sqrt{n}\nonumber\\
U_{\cdot 3} & = \begin{pmatrix} 1 & 1 & -1 & -1 & 1 & 1 & \ldots & -1 & -1  \end{pmatrix}^T/\sqrt{n}.\label{eqn:U}
\end{align}
For $k=1,\ldots,K$, we generated the three diagonal elements of $\Lambda^k$ independently from uniform distributions:   $\Lambda^k_{11} \sim \text{Uniform}(8,15)$, $\Lambda^k_{22} \sim \text{Uniform}(1,4)$ and $\Lambda^k_{33} \sim \text{Uniform}(0,1)$. Note that this choice of parameters results in $U\Lambda^k U^T\in [0,1]^{n\times n}$. We generated data under the model \eqref{eq:mrdpg} with $f$ as the identity, so that $P\left( A_{ij}^k=1\right) = \left( U \Lambda^k U^T\right)_{ij}$.
Under this model, we generated $K \in \{2,4,6,8,10,20,30,40,50\}$ graphs. 
We obtained RDPG$_{\text{separate}}$ by fitting a separate RDPG model to each adjacency matrix, and we obtained RDPG$_{all}$ by fitting a single RDPG model to all $K$ adjacency matrices.

\begin{figure}[h!]
    \centering
    \begin{subfigure}[b]{0.485\textwidth}
        \includegraphics[width=\textwidth]{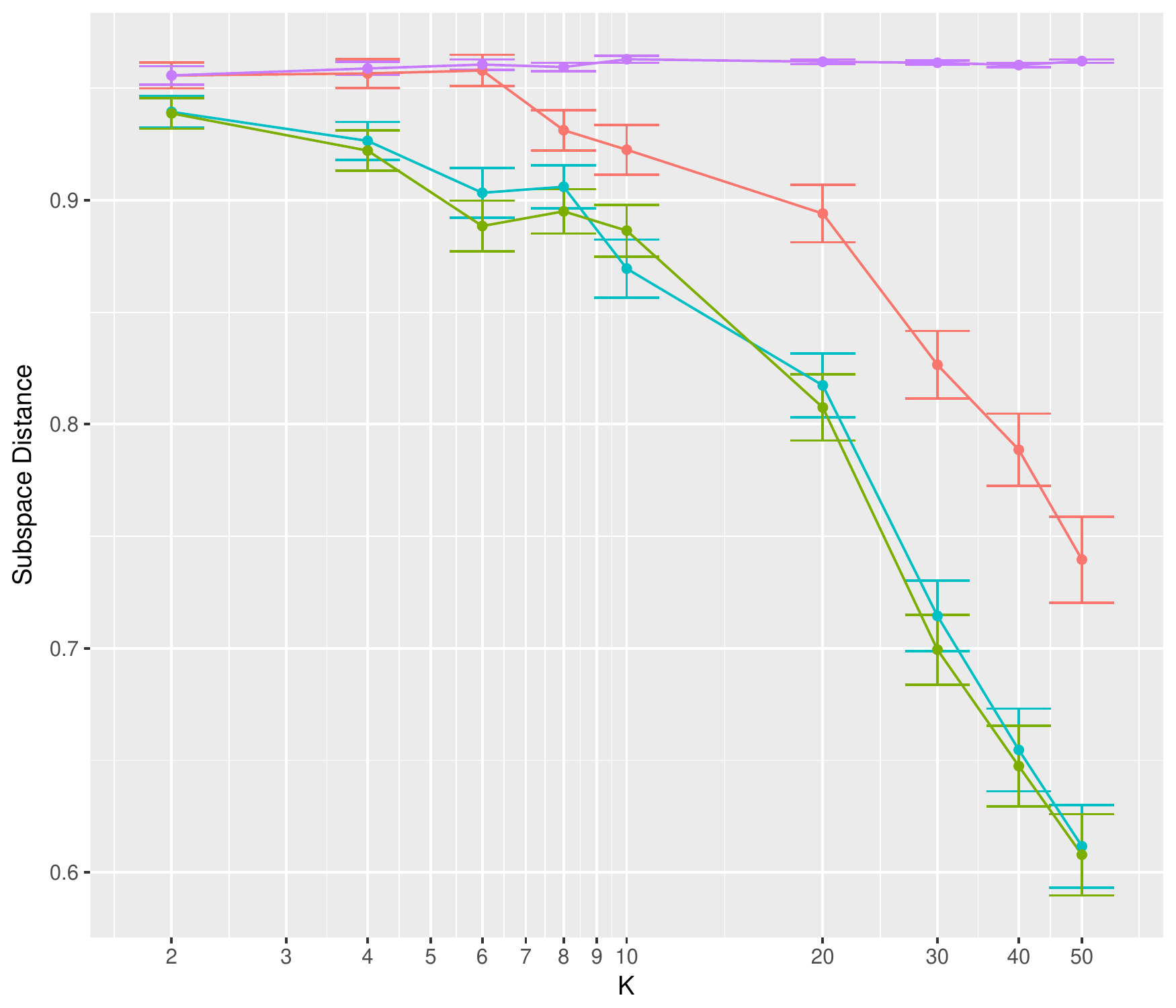}
        \caption{Subspace distance, $d_U$ \eqref{eq:subspacedist}, between the true $U$ and estimated $\hat{U}$.}\label{fig:SE1_subspace}
    \end{subfigure}
    ~ 
    \begin{subfigure}[b]{0.485\textwidth}
        \includegraphics[width=\textwidth]{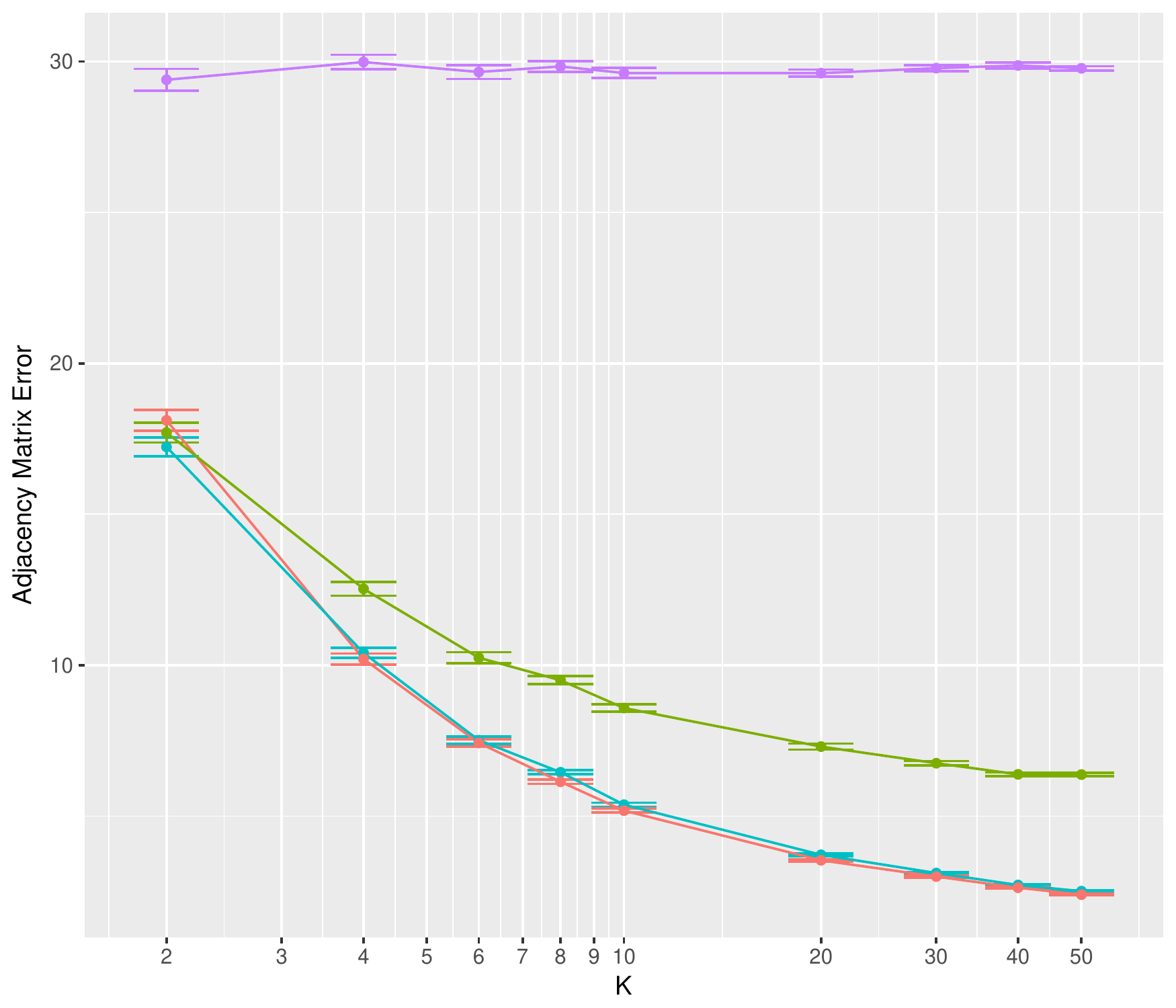}
        \caption{Adjacency matrix error, $d_{\Lambda}$ \eqref{eq:lambdadist}, between the true $\Lambda^1,\ldots,\Lambda^K$ and estimated $\hat{\Lambda}^1,\ldots,\hat{\Lambda}^K$.} \label{fig:SE1_lambda}
    \end{subfigure}
    \caption{Results of multi-RDPG (Algorithm \ref{alg:multiRDPG})  (blue), MREG \citep{wang2017joint} (red), RDPG$_{\text{all}}$ (green), and RDPG$_{\text{separate}}$ (purple), in Simulation Setting 1.  The figures display the mean and standard error, over 100 simulated data sets, of the distance measures defined in \eqref{eq:subspacedist} and \eqref{eq:lambdadist}. }\label{fig:SE1}
\end{figure}

Results, averaged over 100 simulated data sets, are shown in  Figure~\ref{fig:SE1}. Figure \ref{fig:SE1}(a) shows the subspace distance defined in  \eqref{eq:subspacedist}, and Figure \ref{fig:SE1}(b) shows the adjacency matrix error defined in \eqref{eq:lambdadist}. RDPG$_{\text{all}}$ performs the best in terms of subspace distance because it pools adjacency matrices that share the same eigenvectors, but very poorly in terms of adjacency matrix error since it erroneously assumes that $\Lambda^k=\Lambda^{k'}$ for all $k \neq k'$.  RDPG$_{\text{separate}}$ performs poorly using both subspace distance and adjacency error, because it fails to share information across the adjacency matrices. 
Multi-RDPG (Algorithm \ref{alg:multiRDPG}) performs well across the board, and in particular outperforms MREG \citep{wang2017joint} in terms of recovering the matrix $U$.

\subsubsection{Simulation Setting 2}
Next, we expanded Simulation Setting 1 in order to investigate the performance  of our multi-RDPG proposal (Algorithm \ref{alg:multiRDPG}) in a setting where the diagonal elements of $\Lambda^k$ are in different orders for $k=1,\ldots,K$. Once again, we let $n=20$ and $d=3$, with  $U$ defined in \eqref{eqn:U}. For $k$ an even number, we set $\Lambda^k=\Lambda^\mathrm{even} \equiv \text{diag}(11.5,2,0.5)$. For $k$ an odd number, we set $\Lambda^k = \Lambda^{\mathrm{odd}}$, a diagonal matrix for which the diagonal elements are one of six possible permutations of the numbers $11.5$, $2$, and $0.5$. 

We generated each graph according to, 

\begin{equation}\label{eqn:sim2}
P\left( A_{ij}^k=1\right) = f\left( \left(U \Lambda^k U^T\right)_{ij} \right), \quad f(x)=\min(\max(0, x),1).
\end{equation}

For each of the six permutations leading to $\Lambda^\mathrm{odd}$, we generated $K \in  \{2,4,6,8,10,20,30,40,50\}$ graphs. 
Because all of the even-numbered adjacency matrices were drawn from a single generative model, and likewise for the odd-numbered adjacency matrices, we obtained RDPG$_{\text{separate}}$ by fitting one RDPG to all of the even-numbered adjacency matrices, and a separate RDPG to all of the odd-numbered adjacency matrices.

\begin{figure}[!h]
    \centering
    \begin{subfigure}[b]{0.4\textwidth}
      \includegraphics[width=\textwidth]{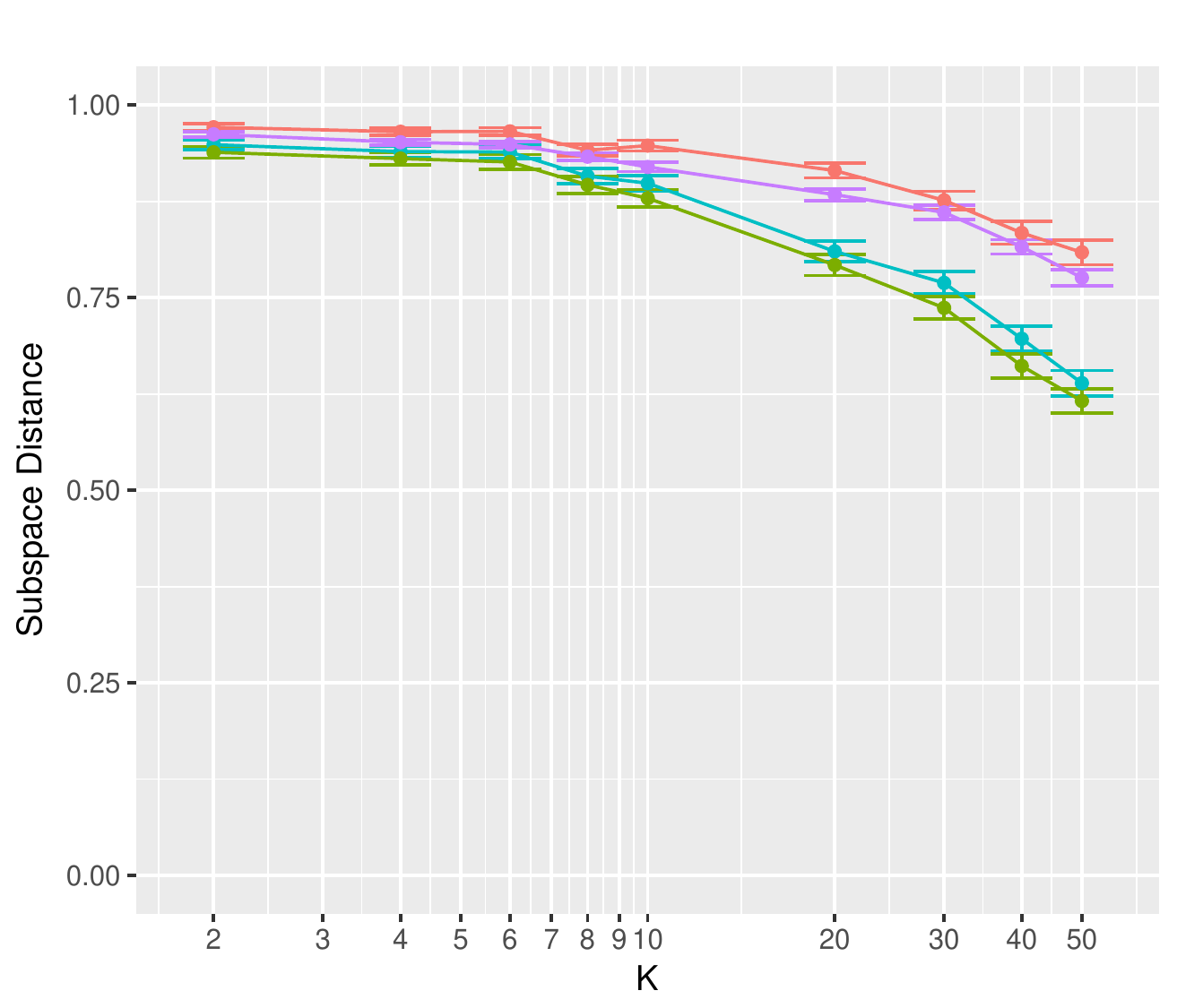}
        \caption{$\Lambda^\mathrm{odd}= \text{diag}(11.5,2,0.5)$}\label{fig:subspace_permutation_a}
    \end{subfigure}
    ~ 
    \begin{subfigure}[b]{0.4\textwidth}
      \includegraphics[width=\textwidth]{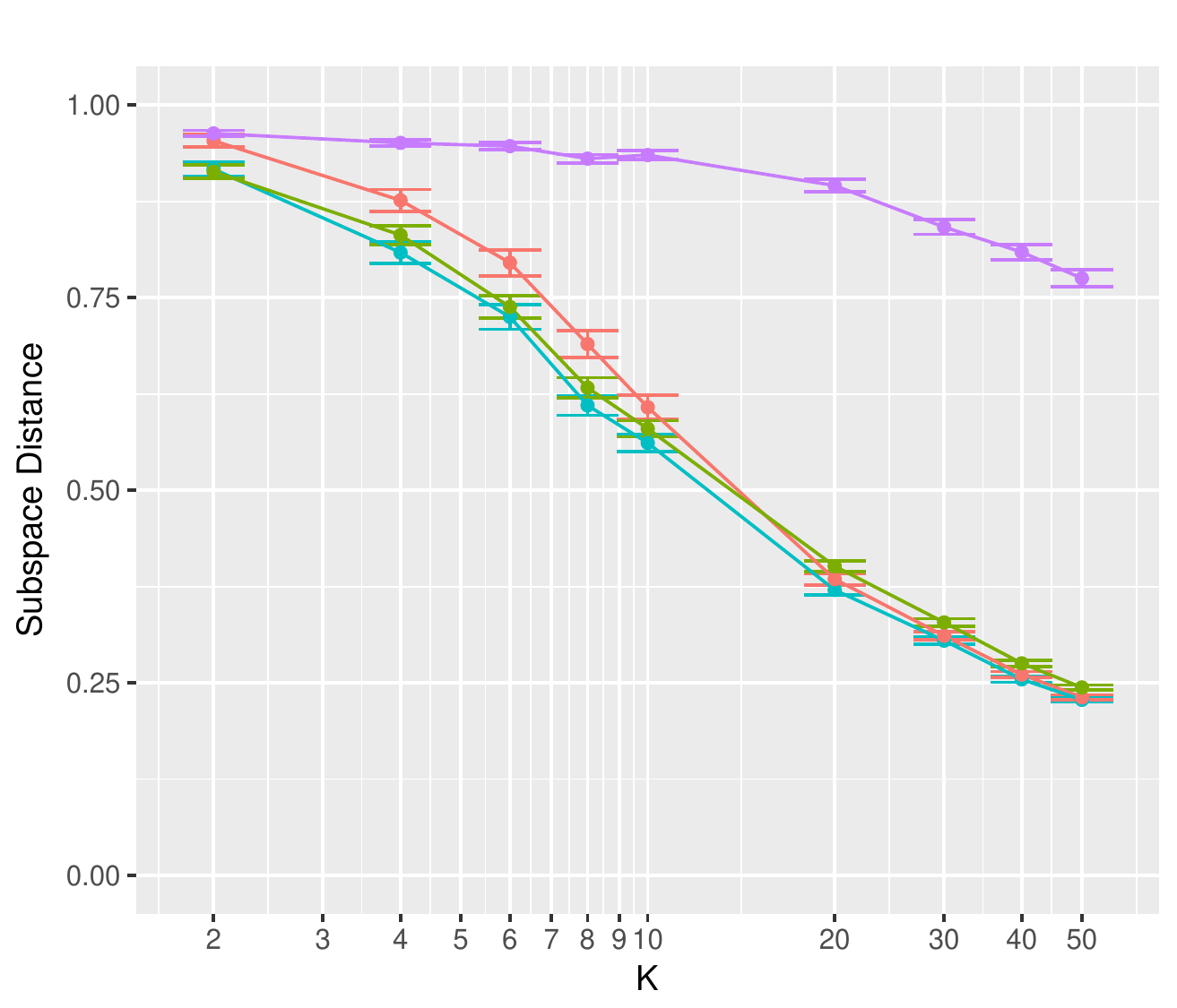}
        \caption{$\Lambda^\mathrm{odd}= \text{diag}(11.5,0.5,2)$}\label{fig:subspace_permutation_b}
    \end{subfigure}
    
        \begin{subfigure}[b]{0.4\textwidth}
      \includegraphics[width=\textwidth]{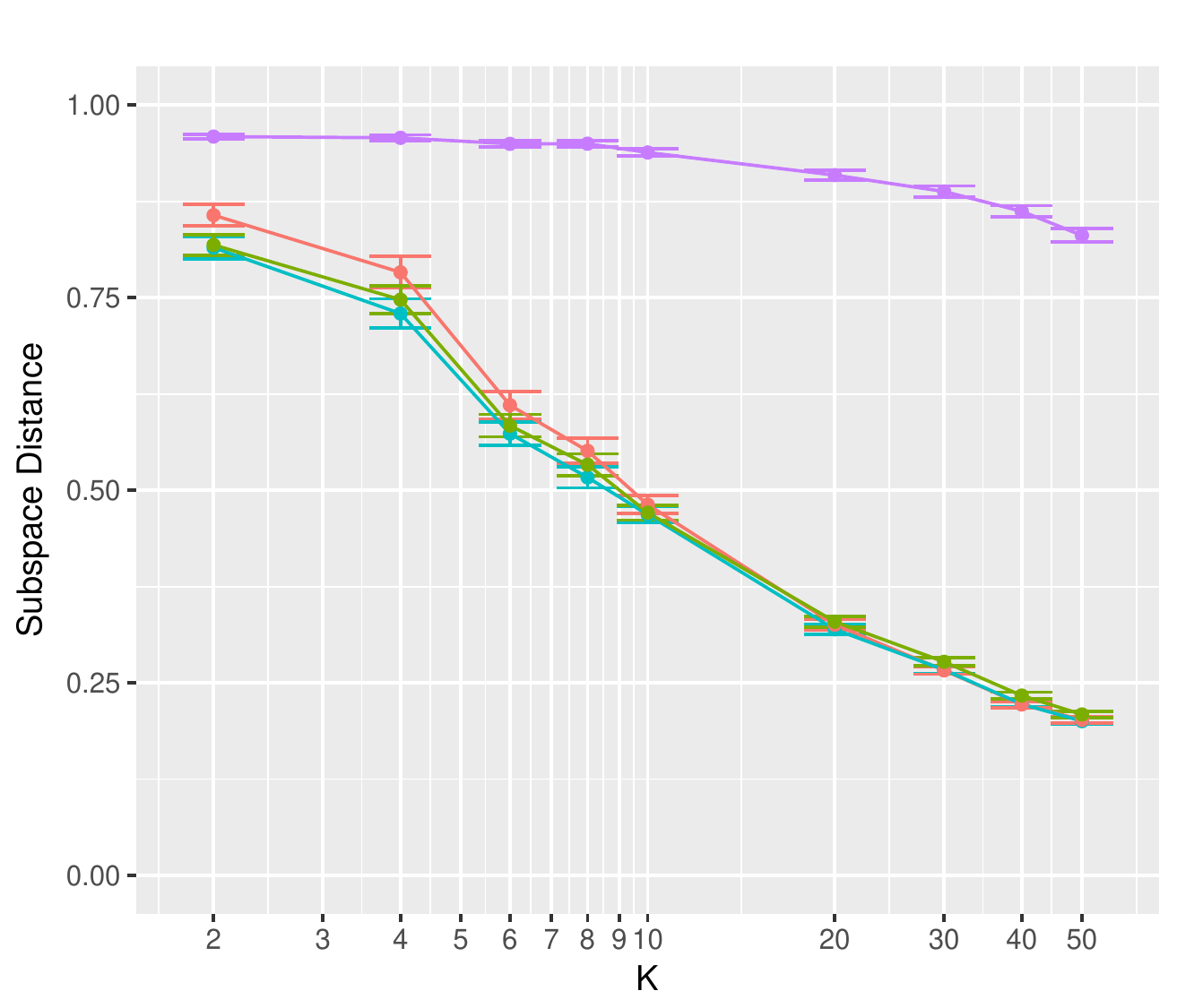}
        \caption{$\Lambda^\mathrm{odd}= \text{diag}(2,0.5,11.5)$}\label{fig:subspace_permutation_c}
    \end{subfigure}
    ~ 
    \begin{subfigure}[b]{0.4\textwidth}
      \includegraphics[width=\textwidth]{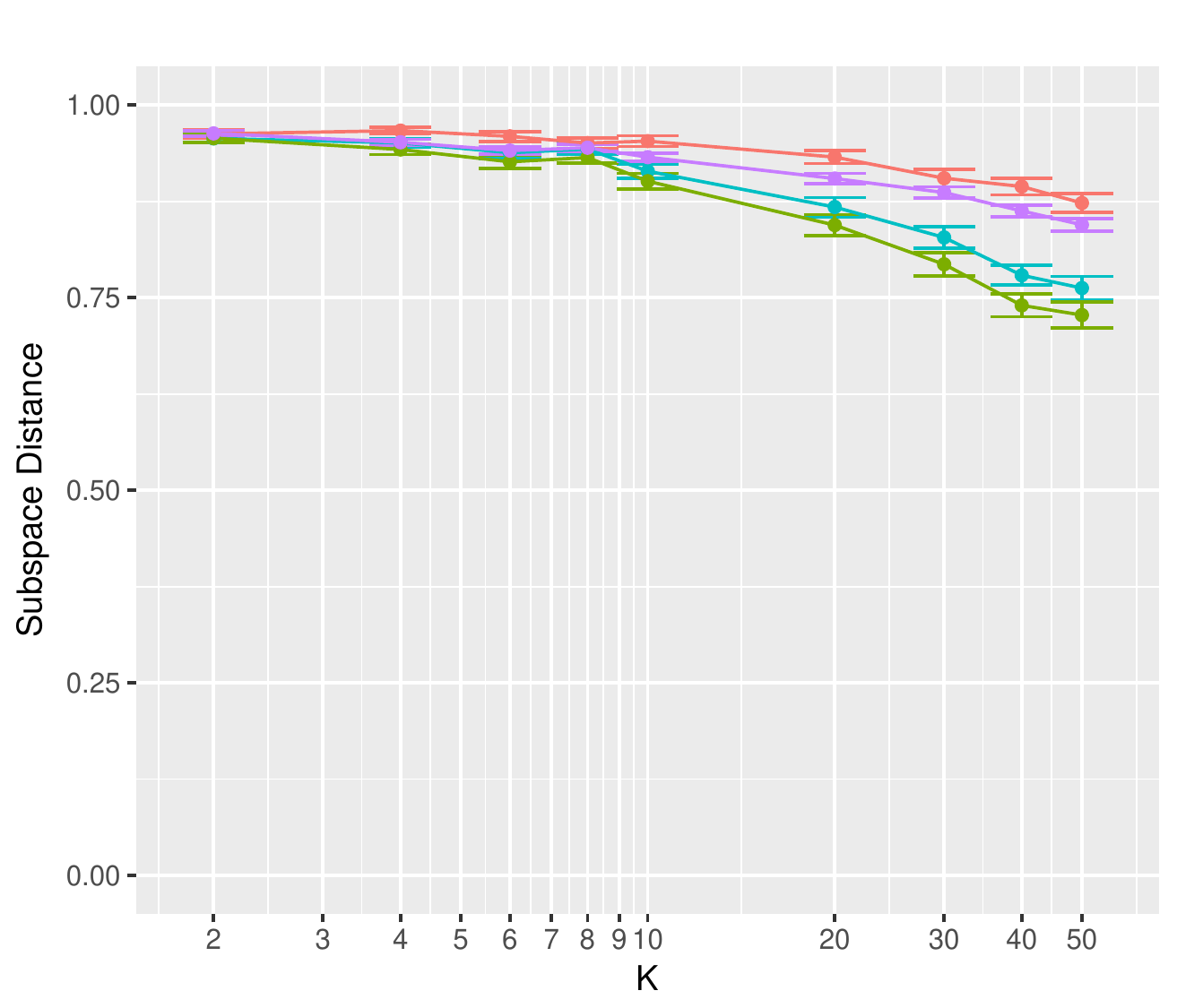}
        \caption{$\Lambda^\mathrm{odd}= \text{diag}(2,11.5,0.5)$}\label{fig:subspace_permutation_d}
    \end{subfigure}
    
        \begin{subfigure}[b]{0.4\textwidth}
      \includegraphics[width=\textwidth]{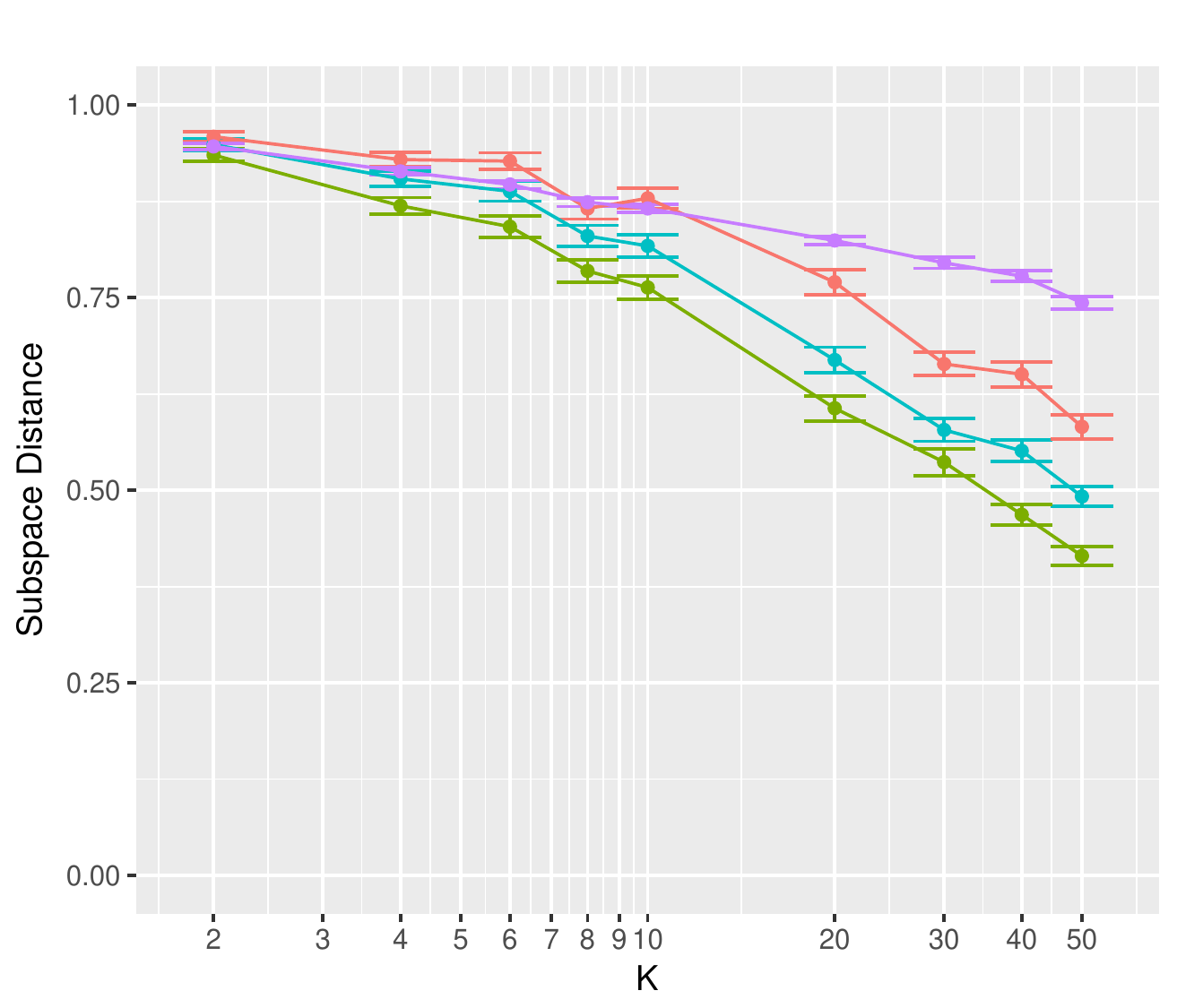}
        \caption{$\Lambda^\mathrm{odd}= \text{diag}(0.5,11.5,2)$}\label{fig:subspace_permutation_e}
    \end{subfigure}
    ~ 
    \begin{subfigure}[b]{0.4\textwidth}
      \includegraphics[width=\textwidth]{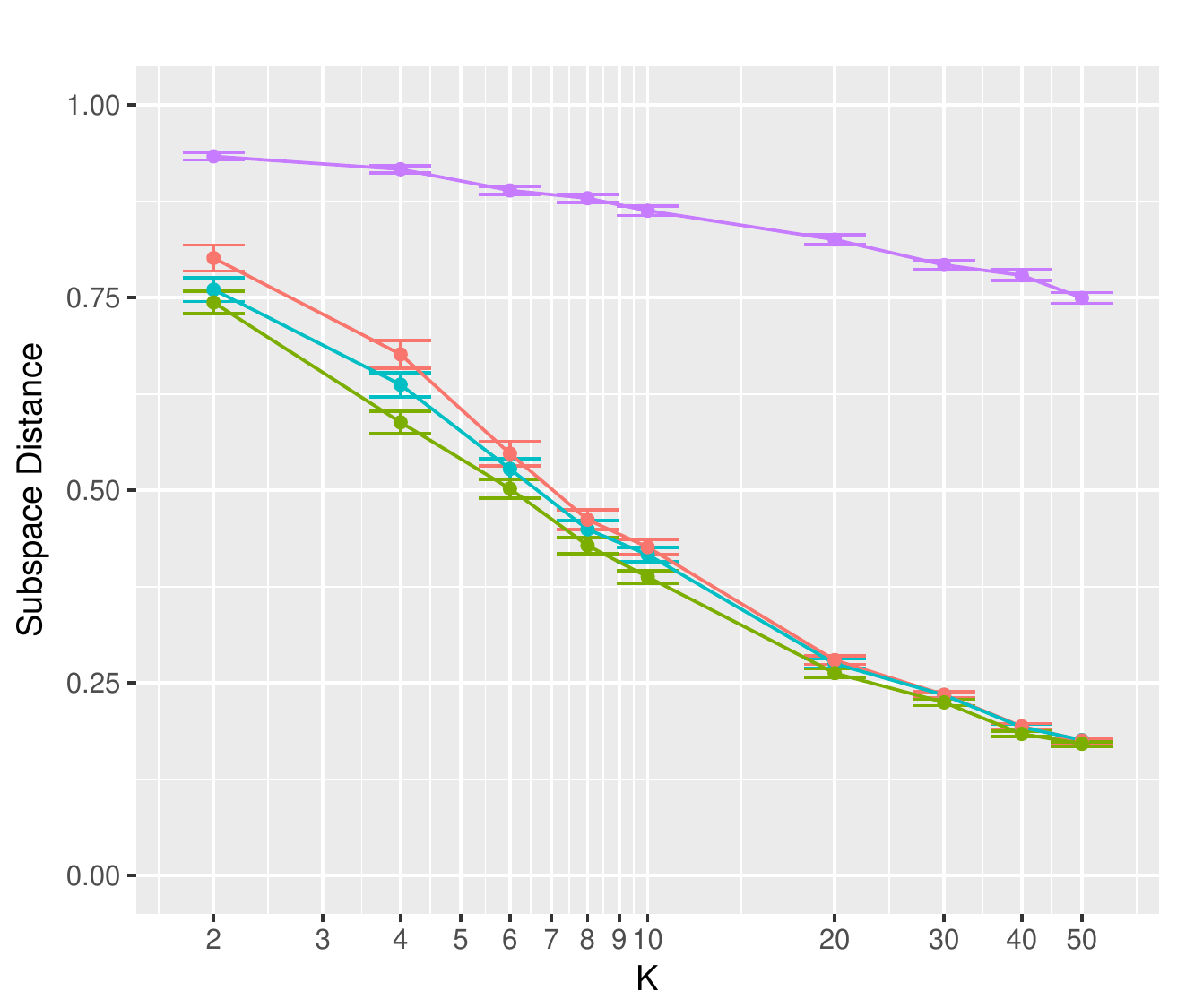}
        \caption{$\Lambda^\mathrm{odd}= \text{diag}(0.5,2,11.5)$}\label{fig:subspace_permutation_f}
    \end{subfigure}
   \caption{Results of multi-RDPG (Algorithm \ref{alg:multiRDPG})  (blue), MREG \citep{wang2017joint} (red), RDPG$_{\text{all}}$ (green), and RDPG$_{\text{separate}}$ (purple), in Simulation Setting 2.  The figures display the mean and standard error, over 100 simulated data sets, of the subspace distance defined in \eqref{eq:subspacedist}. In each subplot, for $k$ even, $\Lambda^k = 
 \text{diag}(11.5, 2, 0.5)$, and for $k$ odd, $\Lambda^k$ is as specified in the subplot caption.}\label{fig:subspace_permutation}
\end{figure}

\begin{figure}[!h]
    \centering
    \begin{subfigure}[b]{0.4\textwidth}
      \includegraphics[width=\textwidth]{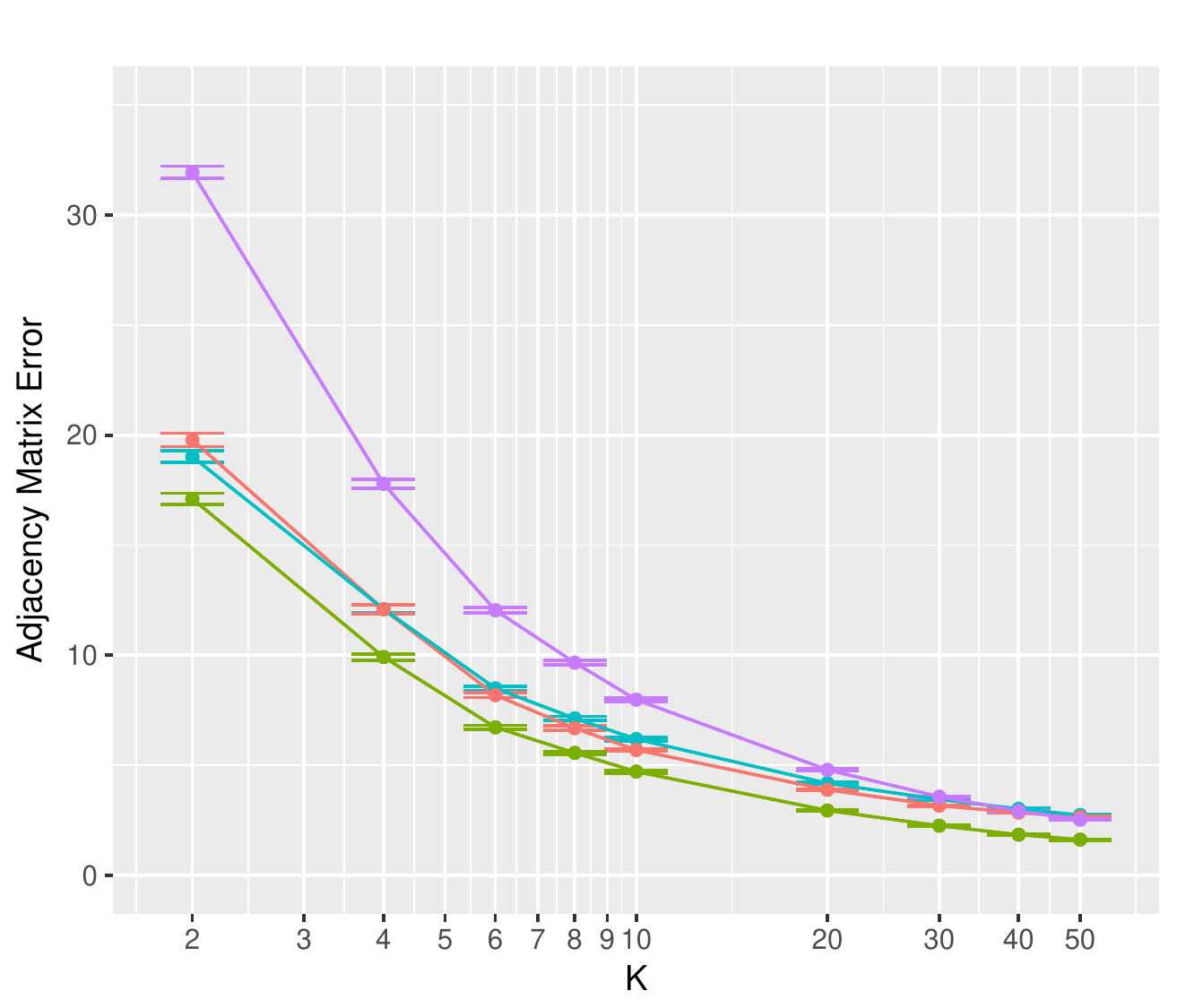}
        \caption{$\Lambda^\mathrm{odd}= \text{diag}(11.5,2,0.5)$}\label{fig:lambda_permutation_a}
    \end{subfigure}
    ~ 
    \begin{subfigure}[b]{0.4\textwidth}
      \includegraphics[width=\textwidth]{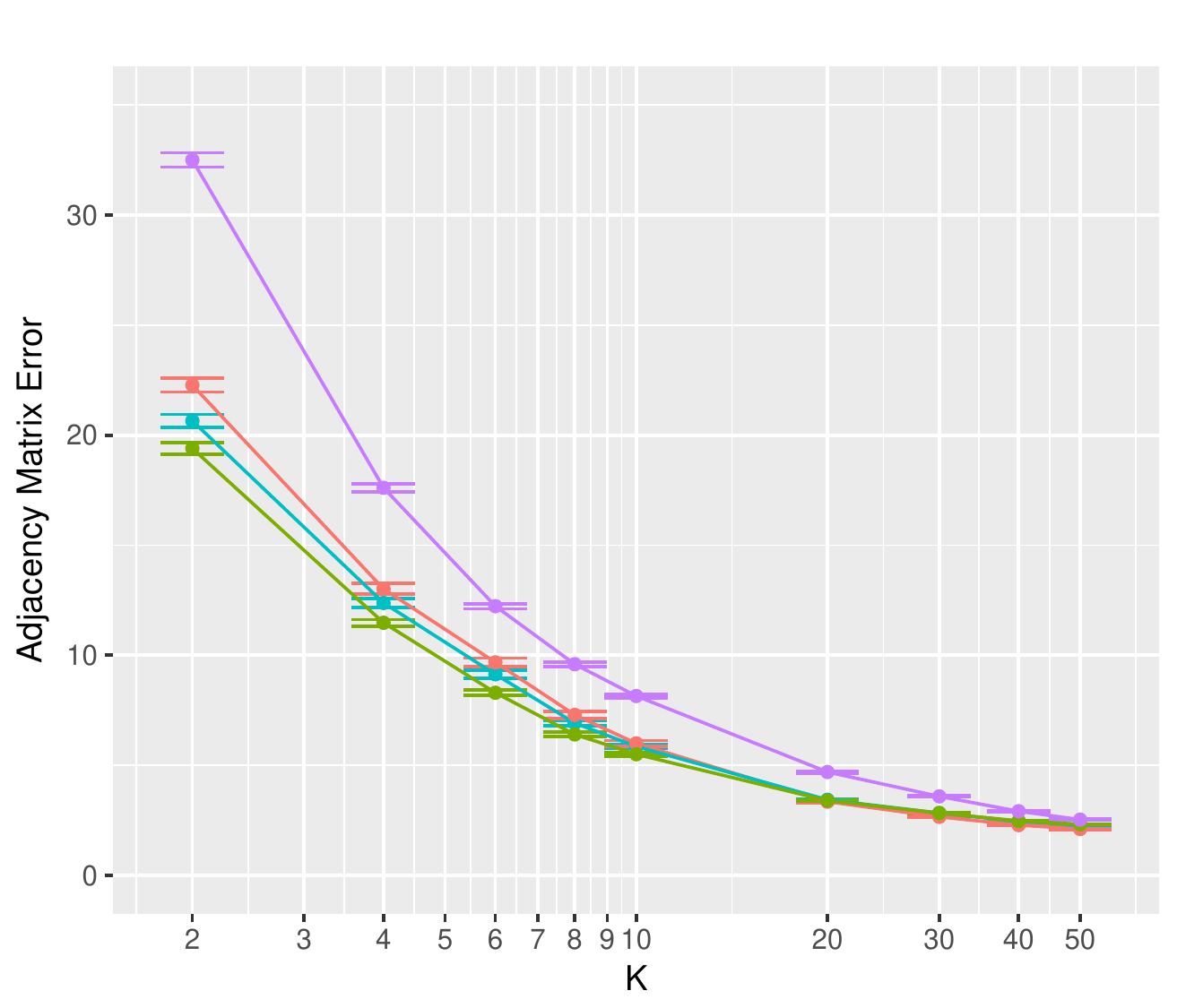}
        \caption{$\Lambda^\mathrm{odd}= \text{diag}(11.5,0.5,2)$}\label{fig:lambda_permutation_b}
    \end{subfigure}
    
        \begin{subfigure}[b]{0.4\textwidth}
      \includegraphics[width=\textwidth]{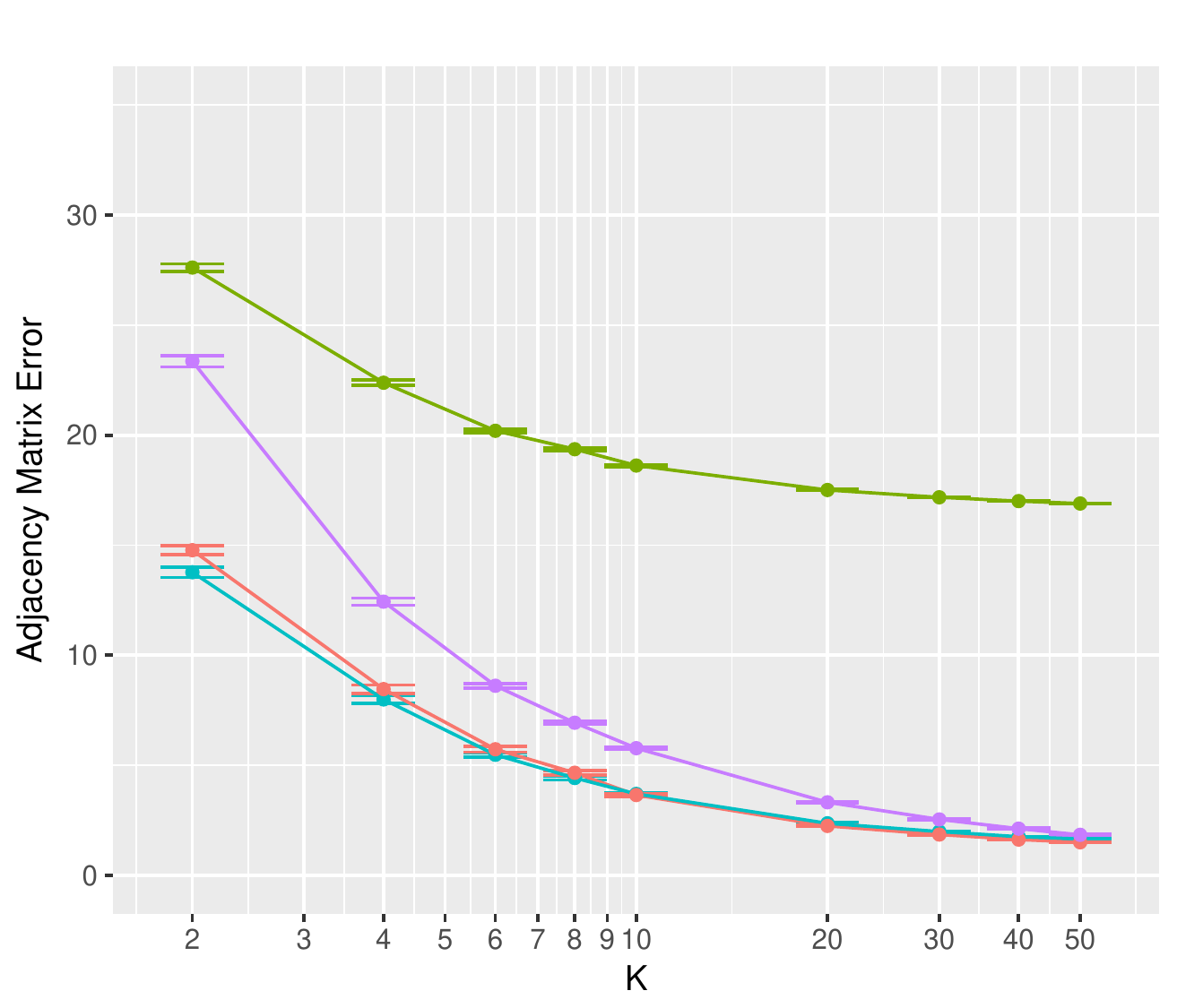}
        \caption{$\Lambda^\mathrm{odd}= \text{diag}(2,0.5,11.5)$}\label{fig:lambda_permutation_c}
    \end{subfigure}
    ~ 
    \begin{subfigure}[b]{0.4\textwidth}
      \includegraphics[width=\textwidth]{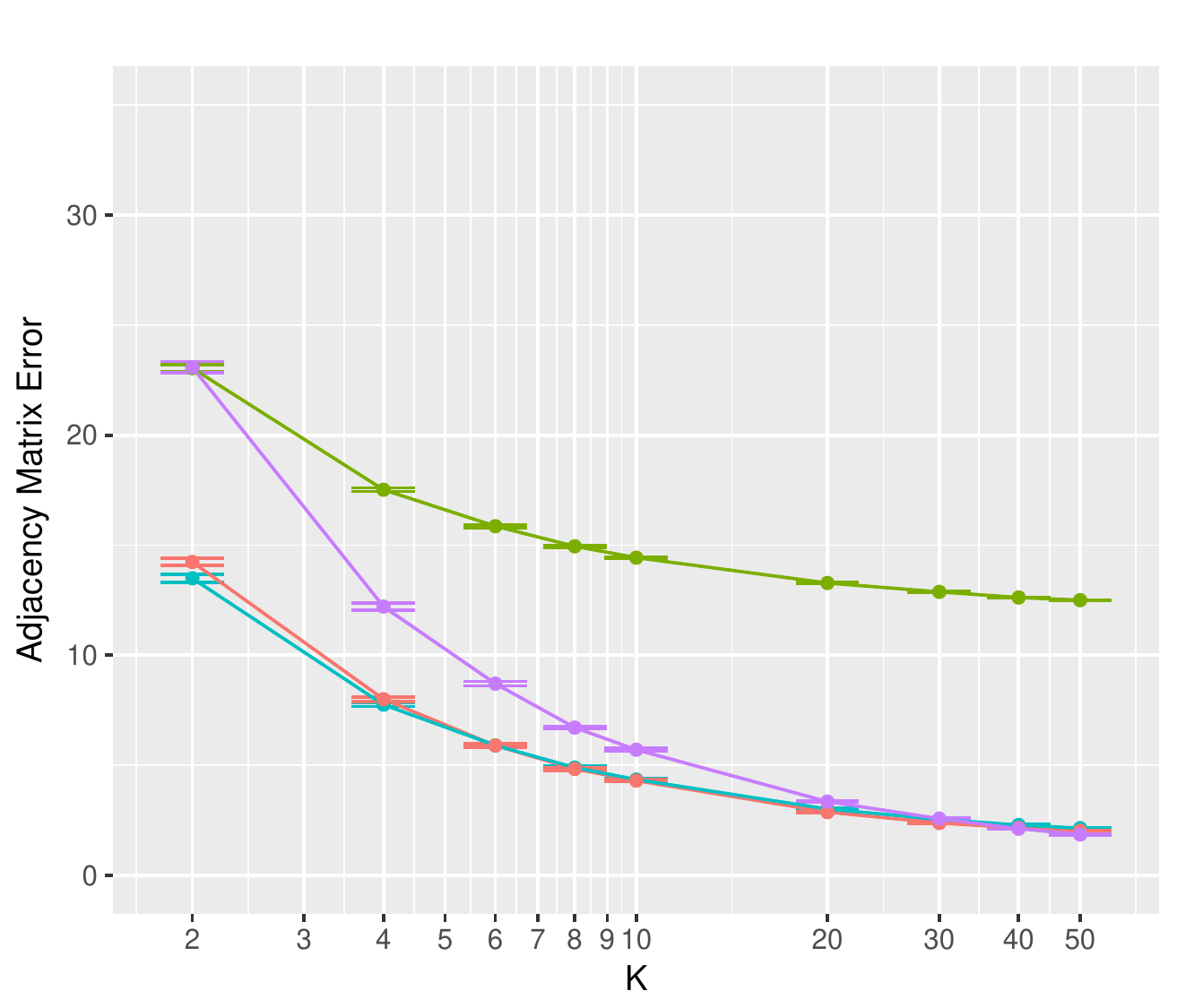}
        \caption{$\Lambda^\mathrm{odd}= \text{diag}(2,11.5,0.5)$}\label{fig:lambda_permutation_d}
    \end{subfigure}
    
        \begin{subfigure}[b]{0.4\textwidth}
      \includegraphics[width=\textwidth]{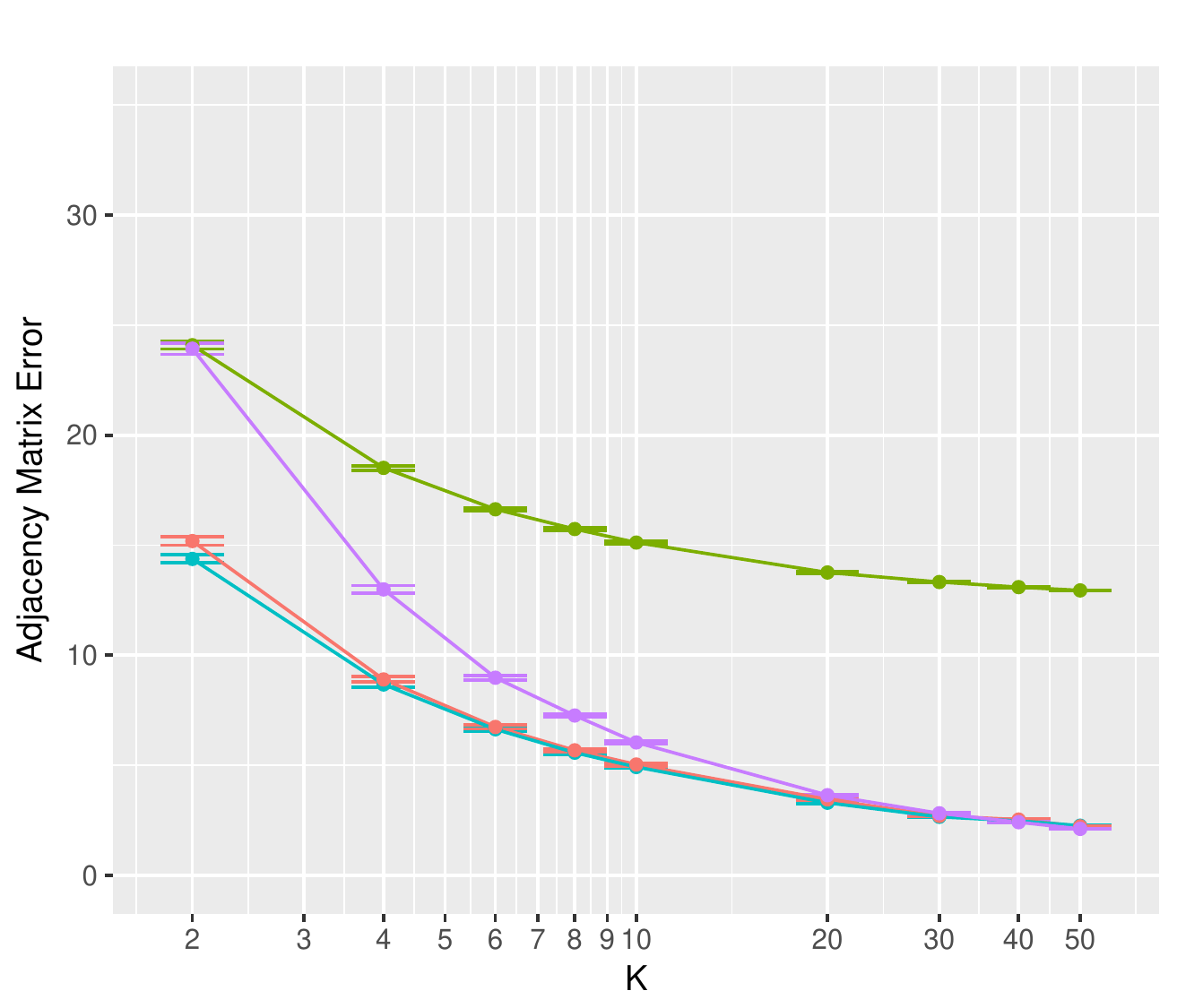}
        \caption{$\Lambda^\mathrm{odd}= \text{diag}(0.5,11.5,2)$}\label{fig:lambda_permutation_e}
    \end{subfigure}
    ~ 
    \begin{subfigure}[b]{0.4\textwidth}
      \includegraphics[width=\textwidth]{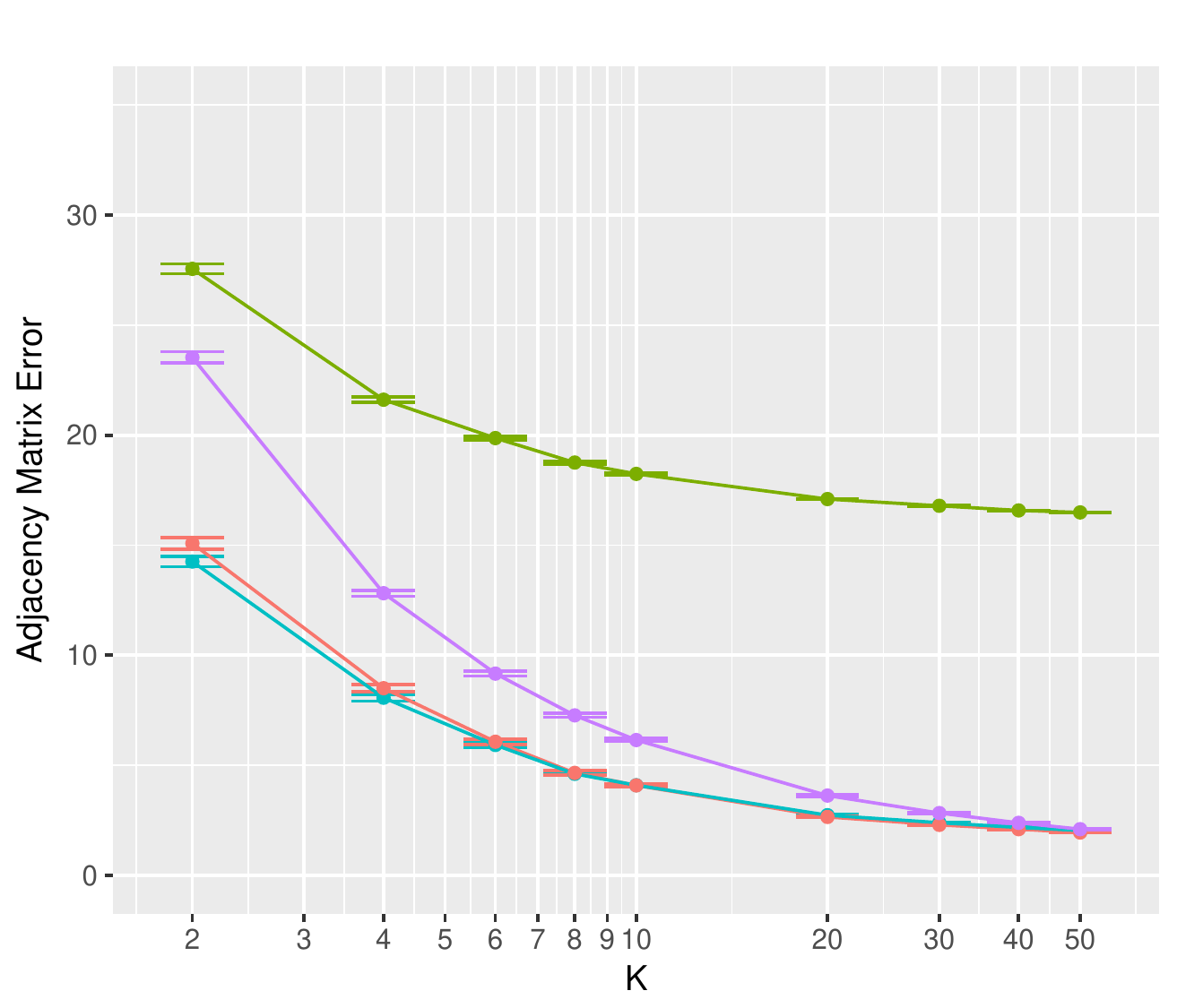}
        \caption{$\Lambda^\mathrm{odd}= \text{diag}(0.5,2,11.5)$}\label{fig:lambda_permutation_f}
    \end{subfigure}
   \caption{Results of multi-RDPG (Algorithm \ref{alg:multiRDPG})  (blue), MREG \citep{wang2017joint} (red), RDPG$_{\text{all}}$ (green), and RDPG$_{\text{separate}}$ (purple), in Simulation Setting 2.  The figures display the mean and standard error, over 100 simulated data sets, of the adjacency matrix error defined in \eqref{eq:lambdadist}. In each subplot, for $k$ even, $\Lambda^k = 
 \text{diag}(11.5, 2, 0.5)$, and for $k$ odd, $\Lambda^k$ is as specified in the subplot caption.}\label{fig:lambda_permutation}
\end{figure}

Results, averaged over 100 simulated data sets, are shown in Figures~\ref{fig:subspace_permutation} and \ref{fig:lambda_permutation}. 
Figure \ref{fig:subspace_permutation} shows the subspace distance defined in \eqref{eq:subspacedist}, and Figure \ref{fig:lambda_permutation} shows the adjacency matrix error defined in  \eqref{eq:lambdadist}.
First, we notice that  RDPG$_{\text{all}}$ always performs well in terms of subspace distance since it pools adjacency matrices that share the same eigenvectors, but quite poorly in terms of adjacency matrix error when the diagonal elements of $\Lambda^k$ differ between the evens and the odds (see, for example, panels (c), (d), (e), and (f) of Figure~\ref{fig:lambda_permutation}). 
The method RDPG$_{\text{separate}}$ performs similarly in all setups, since it fits separate models to the even-numbered and odd-numbered adjacency matrices and borrows no strength across them; overall, its performance is quite poor, because it makes use of only half of the available sample size in fitting each RDPG model. 
We see that multi-RDPG outperforms MREG across the board, using both subspace distance and adjacency matrix error; furthermore,  multi-RDPG has the best overall performance. 

Finally, we notice from Figure~\ref{fig:subspace_permutation} that some of the orderings of the eigenvalues in this simulation setting appear to be more challenging than others. For instance, the errors associated with multi-RDPG in panels (b), (c), (e), and (f) of Figure~\ref{fig:subspace_permutation} are much lower than those in panels (a) and (d). It turns out that panels (a) and (d) are challenging because the eigenvalue corresponding to the third column of $U$ equals 0.5 in both $\Lambda^{\text{even}}$ and $\Lambda^{\text{odd}}$, so that the third column of $U$ is hard to recover. By contrast, the setups in Figures (b), (c), (e), and (f) are less challenging because each of the three eigenvectors of $U$ has an eigenvalue that is no smaller than two in either  $\Lambda^{\text{even}}$ or $\Lambda^{\text{odd}}$.
 Furthermore, the setups in  (d) and (e) are additionally challenging because a substantial proportion of the elements of $U\Lambda^{\text{odd}}U^T$ are less than zero or greater than one; these elements are set to zero by $f(\cdot)$ in \eqref{eqn:sim2}, leading to a substantial loss of information. 

To conclude, we find that while RDPG$_{\text{all}}$ performs well in terms of subspace distance, it typically performs quite poorly in terms of adjacency matrix error, as expected.  RDPG$_{\text{separate}}$ performs poorly because it only makes use of half of the available adjacency matrices. Multi-RDPG has the best overall performance, and substantially outperforms the MREG approach of \cite{wang2017joint}.


\section{A Test for $H_0: \Lambda^1=\ldots=\Lambda^K$} \label{sec:test}

In this section, we will develop a test for the null hypothesis that all of the $K$ observed graphs are drawn from the same distribution; this corresponds to the null hypothesis 
\begin{equation}\label{eqn:nullhypothesis}
H_0: \Lambda^1=\ldots=\Lambda^K
\end{equation}
in the model \eqref{eq:mrdpg}. 
\subsection{A Permutation Test}
To test $H_0: \Lambda^1=\ldots=\Lambda^K$, we take an approach inspired by a likelihood ratio test. 
 The test statistic, $T(A^1_+,\ldots,A^K_+)$, takes the form 
\begin{equation}\label{eqn:Tnull}
\begin{aligned}
T\left(A^1_+,\ldots,A^K_+\right) = & \left( \min_{{U\in \mathbb{R}^{n\times d},\; U^TU=I, \; \Lambda \in \Delta_+^d}} \sum_{k=1}^K ||A^k_+-U\Lambda U^T||_F^2 \right)\\
 &-\left( \min_{{U\in \mathbb{R}^{n\times d},\; U^TU=I, \;  \Lambda^1,\ldots,\Lambda^K \in \Delta_+^d}} \sum_{k=1}^K ||A^k_+-U\Lambda^k U^T||_F^2  \right).
\end{aligned}
\end{equation}
(Recall that $A_+^k$ was defined in Section \ref{sec:algo}.) 
Computing the second term in $T(A_+^1,\ldots,A_+^K)$  is straightforward, using Algorithm~\ref{alg:multiRDPG}. 
To compute the first term, we will make use of the following result. 
\begin{prop}\label{prop:null}
Let $Q D Q^T$ denote the eigen decomposition of $\frac{1}{K} \sum_{k=1}^K A_+^k$, where the diagonal elements of $D = \mathrm{diag}(\alpha_1,\ldots,\alpha_n)$ are in non-increasing order, i.e. $\alpha_1 \geq \alpha_2 \geq \ldots \geq \alpha_n$. 
Then, the solution to 
\begin{equation}
\label{eqn:nullproblem+}
 \minimize_{{U\in \mathbb{R}^{n\times d},U^TU=I,\Lambda \in \Delta_+^d}} \left\{ \sum_{k=1}^K ||A_+^k-U\Lambda U^T||_F^2 \right\}
 \end{equation}
 \\
\noindent is that the $d$ columns of $U$ are the first $d$ columns of $Q$, and the diagonal elements of $\Lambda$ are $\alpha_1,\ldots,\alpha_d$.
\end{prop}

\noindent The proof is provided in  Appendix \ref{app:proofs}. 

We compute a p-value for $H_0: \Lambda^1=\ldots=\Lambda^K$ by comparing the magnitude of $T(A^1_+,\ldots,A^K_+)$ to its null distribution, obtained by permutations. Details are provided in Algorithm~\ref{alg:multiRDPGtest}.
%

%
%
%
\begin{algorithm}
    \caption{A Test for $H_0: \Lambda^1=\ldots=\Lambda^K$ in \eqref{eq:mrdpg}}\label{alg:multiRDPGtest}
            \begin{enumerate}
            \item Compute $T\left(A^1_+,\ldots,A^K_+\right)$ according to \eqref{eqn:Tnull}. 
            \item For $b=1,\ldots,B$:        
		\begin{enumerate}
		\item  Generate $A^{1,*b}_+,\ldots,A^{K,*b}_+$ as follows:
		\begin{enumerate}
		\item For all $i \leq j$, let $\left(A^{1,*b}\right)_{ij},\ldots,\left(A^{K,*b}\right)_{ij}$ be a random permutation of $\left(A^{1}\right)_{ij},\ldots,\left(A^{K}\right)_{ij}$. 
		\item For all $i < j$ and all $k=1,\ldots,K$, set $\left(A^{k,*b}\right)_{ji}$ equal to $\left(A^{k,*b}\right)_{ij}$. 
		\item Let $A^{1,*b}_+,\ldots,A^{K,*b}_+$ be the positive semi-definite parts of $A^{1,*b},\ldots,A^{K,*b}$.
		\end{enumerate}
            		\item  Compute $T\left(A^{1,*b}_+,\ldots,A^{K,*b}_+\right)$ according to \eqref{eqn:Tnull}. 
            	\end{enumerate} 
            	\item Compute the p-value,   $$p = \frac{1}{B} \sum_{b=1}^B I_{\left\{ T\left(A^{1,*b}_+,\ldots,A^{K,*b}_+\right) \geq T\left(A^1_+,\ldots,A^K_+\right)\right\}},$$ where $I_{C}$ is an indicator function that equals one if the event $C$ holds, and equals zero otherwise.
	\end{enumerate}
\end{algorithm}

\subsection{Simulation Study}
We now conduct a simulation study to demonstrate that the p-values for $H_0: \Lambda^1=\ldots=\Lambda^K$ obtained via Algorithm~\ref{alg:multiRDPGtest} are uniformly distributed under the null hypothesis,  and that the test has power under the alternative.

\subsubsection{Type I Error Control}

\begin{figure}[h!]
    \centering
    \begin{subfigure}[b]{0.485\textwidth}
        \includegraphics[width=\textwidth]{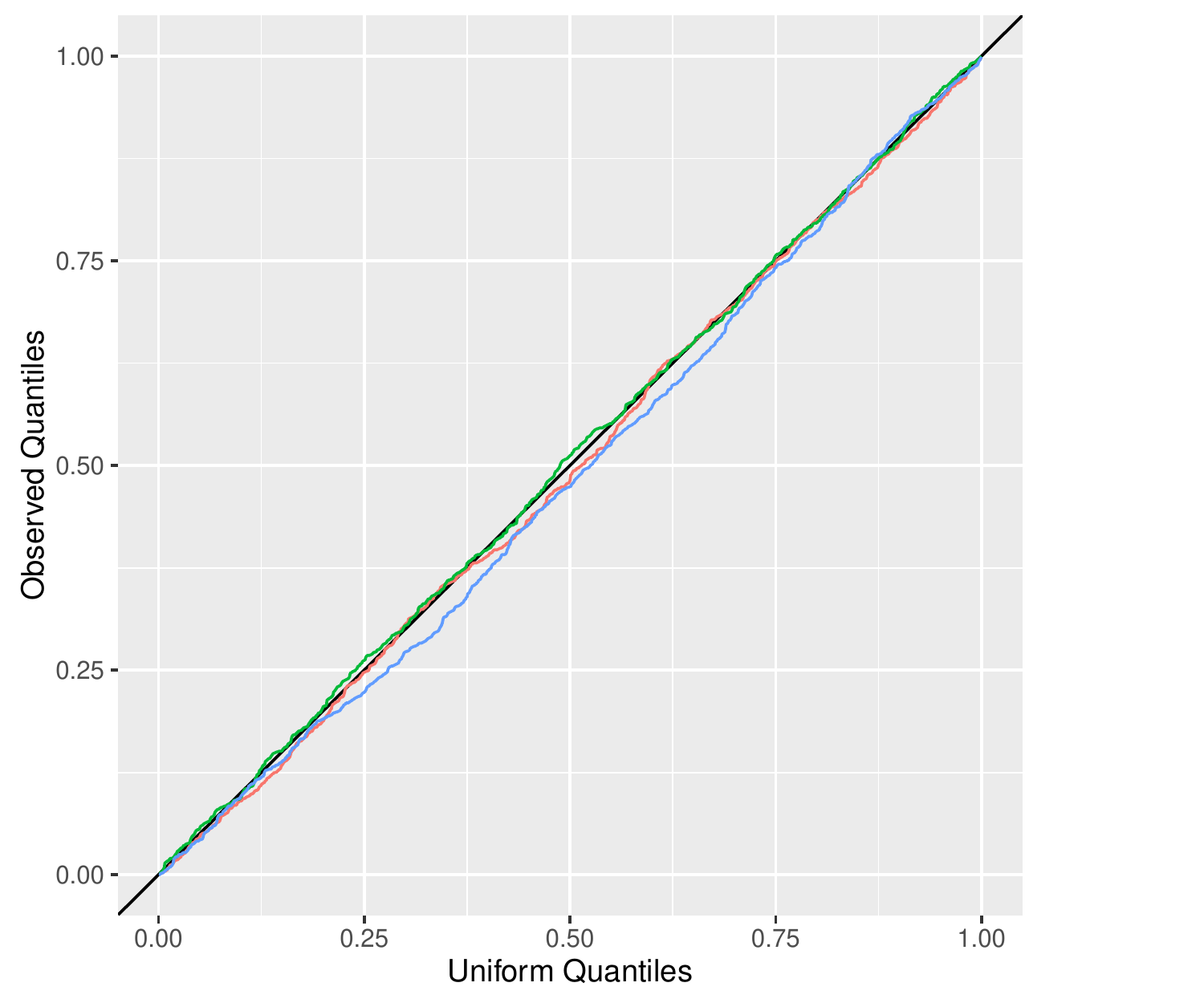}
        \caption{$\Lambda = \text{diag}(n/4,n/5)$}\label{fig:qqplots_n4n5}
    \end{subfigure}
    ~ 
    \begin{subfigure}[b]{0.485\textwidth}
        \includegraphics[width=\textwidth]{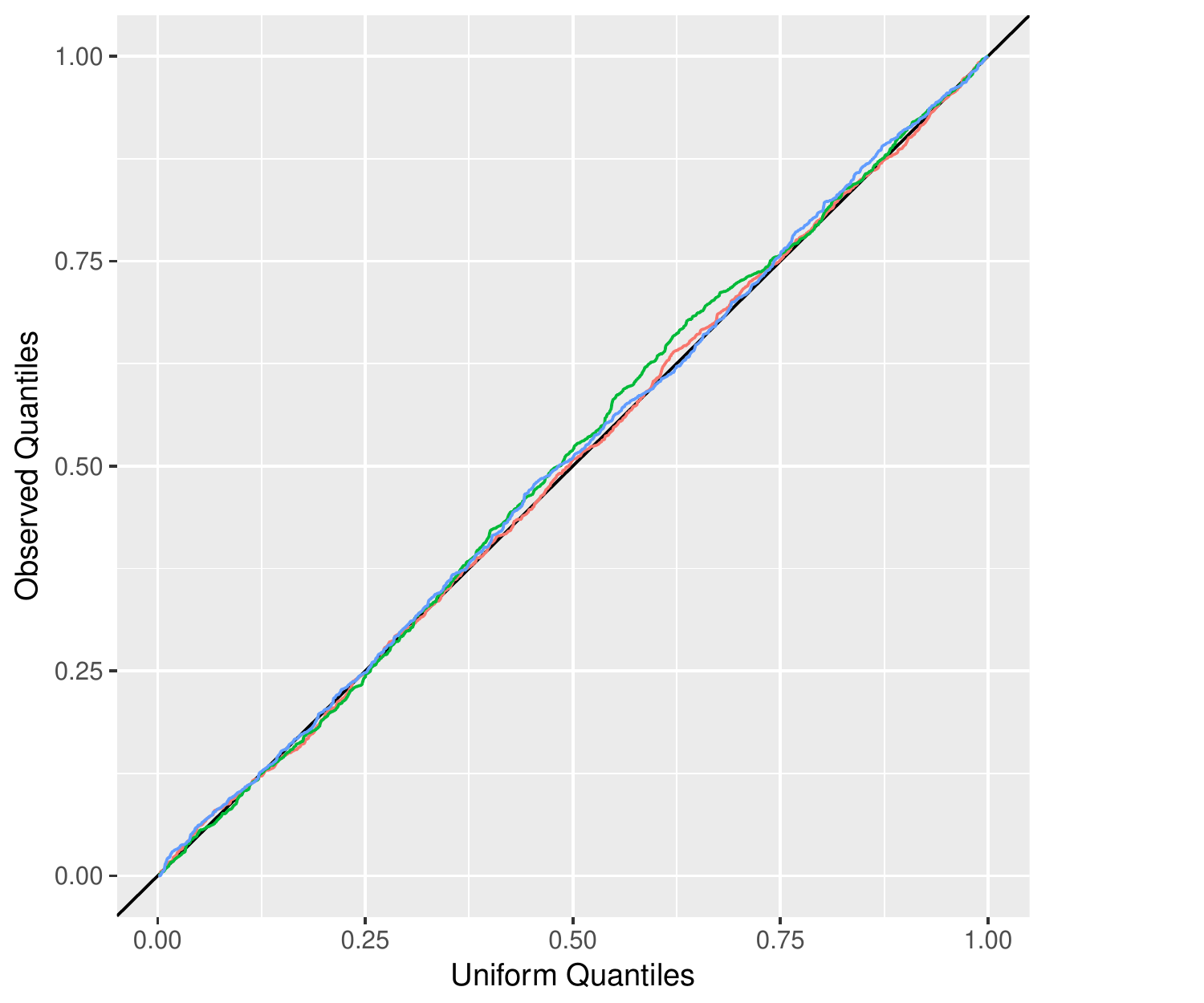}
        \caption{$\Lambda = \text{diag}(n/2,n/4,n/400)$} \label{fig:qqplots_other}
    \end{subfigure}
    \caption{Quantile-quantile plots of p-values from Algorithm~\ref{alg:multiRDPGtest} against a uniform distribution. The colors indicate the value of $n$:  $n=20$ (red), $n=50$ (green), and $n=100$ (blue). \emph{(a):} Data were generated according to \eqref{eq:mrdpg} with $\Lambda^1=\Lambda^2=\text{diag}(n/4,n/5)$ and $U$ as in  \eqref{eqn:U2}. \emph{(b):} Data were generated according to \eqref{eq:mrdpg} with $\Lambda^1=\Lambda^2=\text{diag}(n/2,n/4,n/400)$ and $U$ as in  \eqref{eqn:U}.}\label{fig:qqplots}
\end{figure}

 To explore the Type I error of the test for $H_0: \Lambda^1=\ldots=\Lambda^K$ proposed in Algorithm~\ref{alg:multiRDPGtest}, we generate $K=2$ graphs with 
  \begin{align}\label{eqn:U2}
U  = [[1, \ldots, 1]^T,[1,-1,1,-1,\ldots,-1]^T]/\sqrt{n},
\end{align}
 $\Lambda^1=\Lambda^2=\text{diag}(n/4,n/5)$, for $n\in\{20,50,100\}$.  
For $k=1,2$, we then generated the adjacency matrix $A^k$ according to 
\eqref{eq:mrdpg}, with $f(\cdot)$ the identity. We then computed a p-value according to Algorithm \ref{alg:multiRDPGtest}. 
 We simulated data in this way 1,000 times, and obtained 1,000 p-values, shown in Figure~\ref{fig:qqplots}(a).
 
 We also repeated this procedure using $\Lambda^1=\Lambda^2=\text{diag}(n/2,n/4,n/400)$ and $U$ defined as in \eqref{eqn:U}. The p-values are shown in Figure~\ref{fig:qqplots}(b).
 
In both panels of Figure~\ref{fig:qqplots}, we see that the p-values are uniformly distributed, indicating adequate control of Type I error.

\subsubsection{Power} 

Next, in order to explore the power of the test described in Algorithm~\ref{alg:multiRDPGtest}, we generated data for which $H_0: \Lambda^1 = \ldots=\Lambda^K$ does not hold. 

\begin{figure}[h!]
    \centering
    \begin{subfigure}[b]{0.485\textwidth}
        \includegraphics[width=\textwidth]{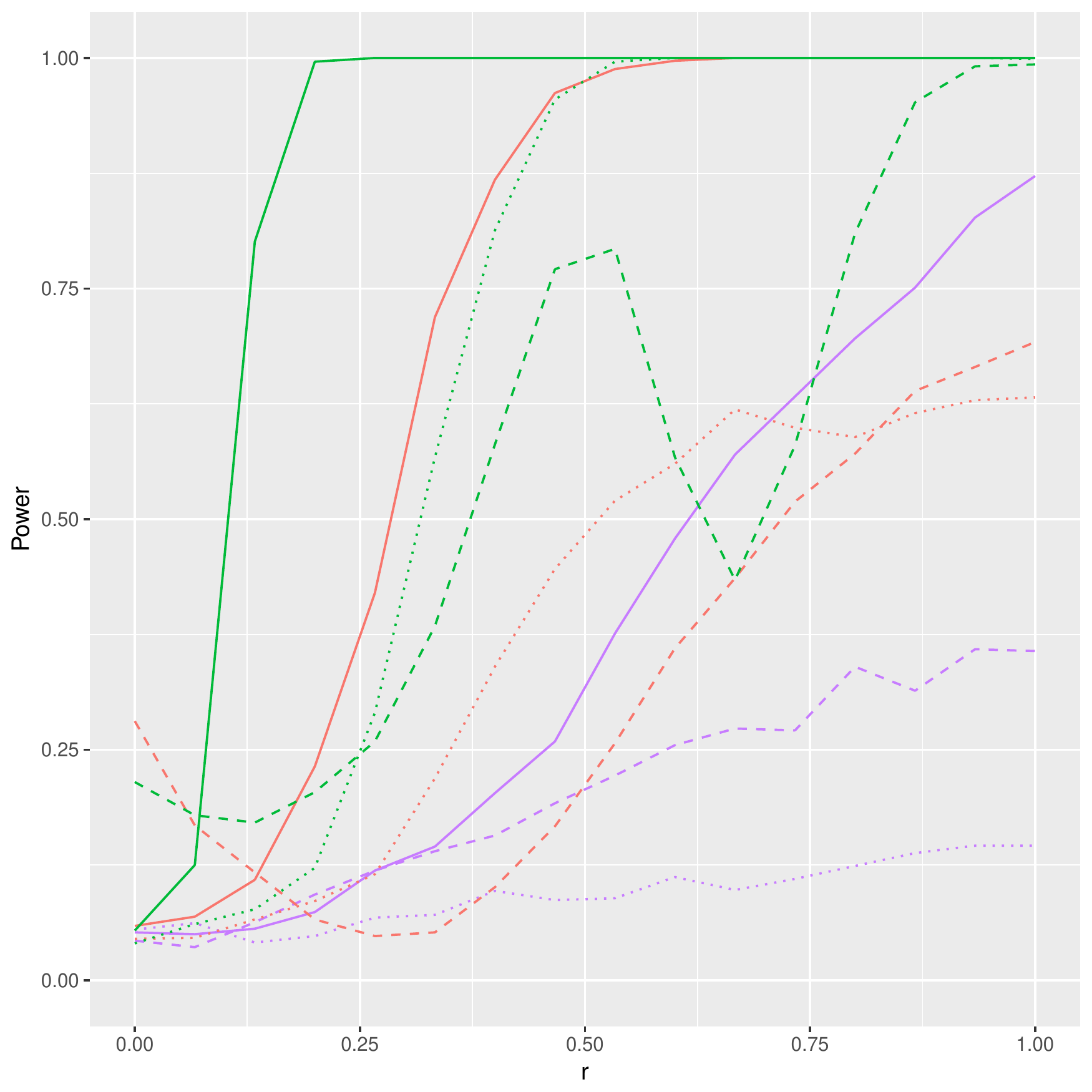}
        \caption{ $\Lambda^1,\Lambda^2$  in \eqref{eq:K2}, and $U$  in  \eqref{eqn:U2}}\label{fig:powerplots_n4}
    \end{subfigure}
    ~ 
    \begin{subfigure}[b]{0.485\textwidth}
        \includegraphics[width=\textwidth]{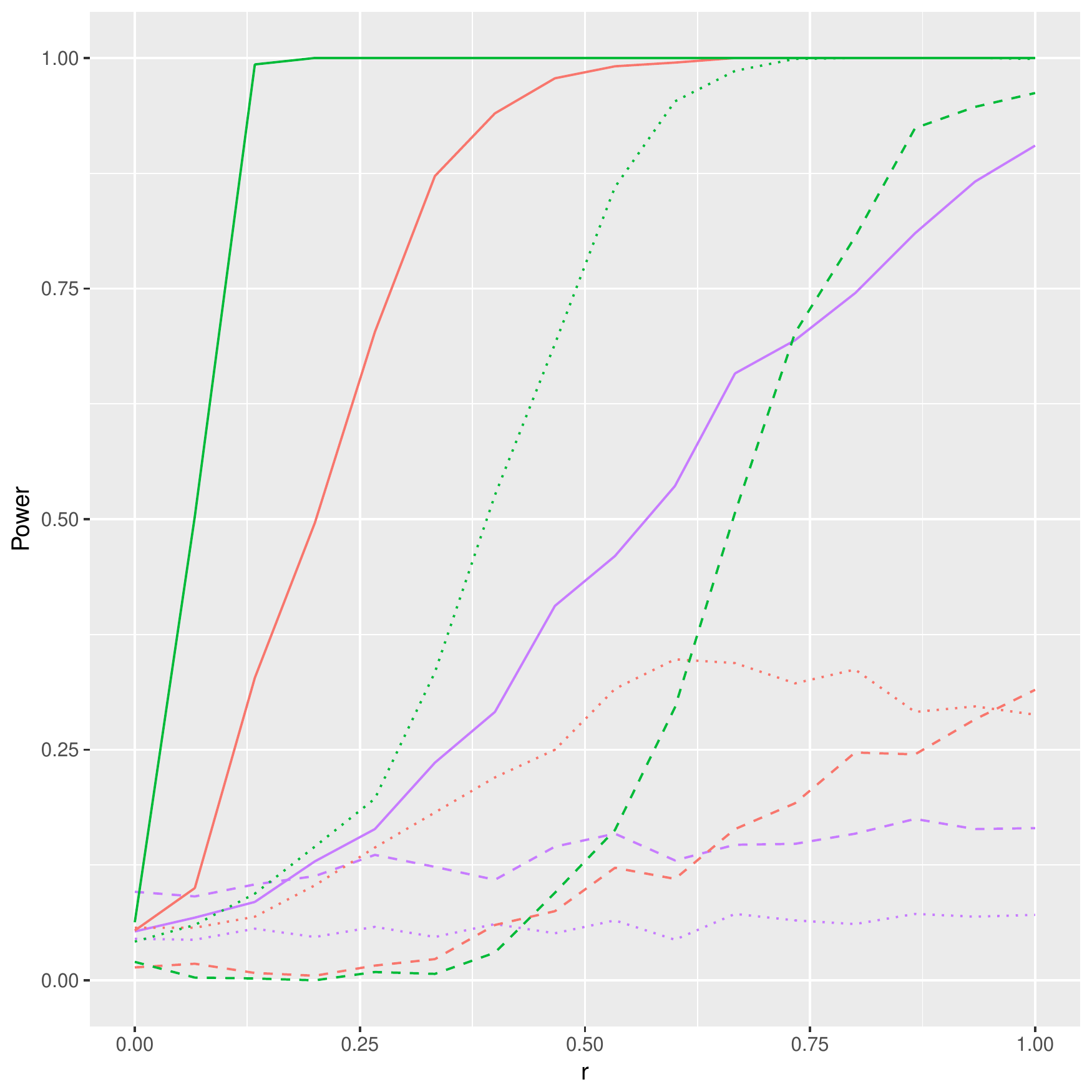}
        \caption{$\Lambda^1,\Lambda^2$  in \eqref{eq:K3}, and $U$  in  \eqref{eqn:U}} \label{fig:powerplots_other}
    \end{subfigure}
    \caption{The power of the test in Algorithm~\ref{alg:multiRDPGtest} (solid), along with the powers of the test of \cite{tang2016semiparametric} (dash) and the test of \cite{tang2016semiparametric} with a modification to the bootstrap method (dotted).   Colors indicate the value of $n$: $n=10$ (purple), $n=20$ (red), and $n=50$ (green).
    \emph{(a)}: $\Lambda^1$ and $\Lambda^2$ are defined in \eqref{eq:K2}, and $U$ is defined in  \eqref{eqn:U2}. \emph{(b):} $\Lambda^1$ and $\Lambda^2$ are defined in \eqref{eq:K3}, and $U$ is defined  in \eqref{eqn:U}.}\label{fig:powerplots}
\end{figure}

First, for $r \in [0,1]$, and for  $n\in \{20,50,100\}$, we generated $\Lambda^1$ and $\Lambda^2$ as follows:
\begin{equation}
\Lambda^1 =  \text{diag}\left(\frac{n}{4} (1+r),\frac{n}{4}(1-r)\right), \quad \Lambda^2 =  \text{diag}\left(\frac{n}{4} (1-r),\frac{n}{4}(1+r)\right). \label{eq:K2}
\end{equation}
%
We generated $U$ as in  \eqref{eqn:U2}. 
We then generated $A^1$ and $A^2$ according to \eqref{eq:mrdpg}, with $f(u)=\max(u,0)$.  A p-value for $H_0: \Lambda^1=\Lambda^2$ was then obtained using Algorithm~\ref{alg:multiRDPGtest}.
This was repeated 1,000 times, leading to p-values $p_1,\ldots,p_{1000}$. The power at level $\alpha=0.05$, computed as
 \begin{equation}
\text{power} = \frac{1}{1000}\sum_{i=1}^{1000}I_{(p_i < \alpha)},
\end{equation}
where $I_C$ is an indicator function for the event $C$,
is displayed in Figure~\ref{fig:powerplots}(a).
As expected, the power increases as $r$ increases, and as $n$ increases.

Next, we defined $U$  as in equation \eqref{eqn:U} with
\begin{equation}\label{eq:K3}
\begin{aligned}
\Lambda^1 =  \text{diag}\left(\frac{n}{4} (1-r),\frac{n}{5} (1+r),\frac{n}{400}(1-r)\right), \\ \Lambda^2 =  \text{diag}\left(\frac{n}{4} (1+r),\frac{n}{5} (1-r),\frac{n}{400}(1+r)\right).
\end{aligned}
\end{equation}
The resulting power is shown in Figure~\ref{fig:powerplots}(b); we see once again that power increases as $r$ and $n$ increase. 
%

%
%

We now compare our test to a test proposed by \cite{tang2016semiparametric}. Under the assumption that both graphs are drawn from a RDPG model \eqref{eq:rdpg}, \cite{tang2016semiparametric} fits the RDPG to each graph, in order to obtain $\hat X^1 \equiv \begin{pmatrix} \hat x_1^1 & \ldots & \hat x_n^1 \end{pmatrix}^T $ and $\hat X^2 \equiv \begin{pmatrix} \hat{x}_1^2 & \ldots & \hat x_n^2 \end{pmatrix}^T $, the two sets of estimated latent vectors. They then calculate the test statistic 
%
$T = \min_W \| \hat{X}^1-\hat{X}^2 W \|_F$, and also calculate the values of this test statistic under a null distribution obtained via the parametric bootstrap. This is then used to obtain a p-value for the null hypothesis that the two graphs are drawn from the same RDPG model. We fit the model using software obtained from the authors of \cite{tang2016semiparametric}.


We see from Figure \ref{fig:powerplots} that with data generated under \eqref{eq:K2} and \eqref{eq:K3}, our test has higher power than that of \cite{tang2016semiparametric}. Under \eqref{eq:K2}, the proposal of \cite{tang2016semiparametric} fails to control Type I error: it rejects the null hypothesis in approximately 20\% of simulated data sets with $n=50$ for which $H_0: \Lambda^1=\Lambda^2$ holds. A slight modification to their algorithm that replaces the parametric bootstrap with the non-parametric swapping approach used in Algorithm \ref{alg:multiRDPGtest} leads to proper Type I error control, but lower power than our approach.

\section{Application to Data}\label{sec:realdata}
\subsection{Wikipedia data}
To begin, we consider graphs representing a subset of Wikipedia \citep{suwan2016empirical}. The data set is accessible at \url{http://www.cis.jhu.edu/~parky/Data/data.html}. The 1,383 vertices represent Wikipedia pages that are available in both French and English. 
An English graph is constructed by placing an edge between a pair of vertices if the English version of either page hyperlinks to the other; the French graph is constructed analogously. 
 We wish to test whether the two graphs are drawn from the same distribution, i.e. we wish to test the null hypothesis $H_0: \Lambda^\text{English}=\Lambda^\text{French}$  in the model \eqref{eq:mrdpg}. 

The English graph contains 18,857 edges, whereas the French graph contains only 14,973 edges. Therefore, we randomly down-sample the edges in the English graph so that both graphs contain 14,973 edges.  We then test $H_0: \Lambda^\text{English}=\Lambda^\text{French}$ using Algorithm \ref{alg:multiRDPGtest}. This results in a test statistic value of $T(A^\text{English}_+,A^\text{French}_+) = 93.08$ and a p-value of $p=0$. Therefore, we reject the null hypothesis that the two graphs come from the same distribution.

As a point of comparison, we also construct two graphs by  randomly sampling 15,000 edges from the English graph twice. This gives two largely overlapping graphs. We test whether these graphs are drawn from the same distribution using Algorithm \ref{alg:multiRDPGtest}. This results in a test statistic value of $T(A^\text{Sample\, 1}_+,A^\text{Sample\, 2}_+) = 0.08$, and a p-value of $p = 0.81$. Therefore, as expected, we fail to reject the null hypothesis that the two graphs come from the same distribution. 

\subsection{\emph{C. elegans} Data}

We now consider brain networks  in \emph{C. elegans},  a small transparent roundworm \citep{varshney2011structural,chen2016joint}. The data set is accessible at \url{https://neurodata.io/project/connectomes/}. The 253 vertices represent neurons. We consider two graphs: a chemical graph, in which two vertices are connected by an edge if there is a chemical synapse between them, and an electrical graph, in which 
 two vertices are connected by an edge if there is an electrical junction potential between them.  We wish to test whether the chemical graph and the electrical graph come from the same distribution. Because the chemical graph has 1695 edges and the electrical graph has only 517 edges, we randomly down-sampled the edges in the former to obtain a graph with only 517 edges.

We tested the null hypothesis $H_0: \Lambda^\text{Chemical}=\Lambda^\text{Electrical}$
 using Algorithm \ref{alg:multiRDPGtest}. This yields a test statistic of $T(A^\text{Electrical}_+,A^\text{Chemical}_+) =18.06$ and a  p-value of $0$. This leads us to reject the null hypothesis. 
\section{Discussion}\label{sec:disc}

In this paper, we propose the multi-RDPG model, a direct extension of the RDPG model  to the setting of multiple graphs. This new model is closely related to the MREG model of  \cite{wang2017joint}. Unlike the MREG model, the multi-RDPG requires the latent vectors to be orthogonal and the corresponding values to be non-negative, so that the multi-RDPG is a direct generalization of the RDPG. Furthermore, we propose a procedure for fitting the multi-RDPG model that allows us to estimate all latent vectors simultaneously, leading to improved empirical results. Finally, we present an approach for testing whether the eigenvalues are equal across the graphs, which controls type I error and has adequate power against the alternative.

In this paper, we have taken the link function $f(\cdot)$  in \eqref{eq:mrdpg} to be the identity, in the interest of simplicity. However, it would be more natural to choose $f(\cdot)$ to be a function that maps from $\mathbb{R}$ to $[0,1]$, such as the logistic function, $f(x)=\exp(x)/(1+\exp(x))$. This would require only a modest modification to Algorithm~\ref{alg:multiRDPG}. We leave the details to future work. 

The R package \verb=multiRDPG= will be posted on CRAN.

\section*{Acknowledgments}

We thank Shangsi Wang and Minh Tang at Johns Hopkins University for providing software implementing the methods in \cite{wang2017joint} and \cite{tang2016semiparametric}, respectively. We thank Debarghya Ghoshdastidar at Eberhard Karls University of T\"ubingen for helpful answers to our questions. D.W. was partially supported by NIH Grants  DP5OD009145 and R01GM123993, and NSF CAREER Award DMS-1252624.  

\appendix
\section{Proofs} \label{app:proofs}

\subsection*{Proof of Proposition~\ref{prop:majorize}.}
\begin{proof}
Lieb's Concavity Theorem \citep{lieb1973convex} states that $\text{trace}(K^T L^p KB^r)$ is convex in $K$ if $L$ and $B$ are positive semi-definite and $p,r\geq 0 $ and $p+r \leq 1$. Applying Lieb's Concavity Theorem with $p=r=0.5$, $K=U^T$, $L=\left( \Lambda^k \right)^2$, and $B=\left(A_+^k\right)^2$, we  see that $\mathrm{trace}\left(U \Lambda^k U^T A_+^k \right)$ is convex in $U$. Therefore, 
$g(U) \equiv -2 \sum_{k=1}^K \mathrm{trace}\left(U \Lambda^k U^T A_+^k \right)$ is concave in $U$. 

By definition of concavity \citep{boyd2004convex}, 
$$g(U) \leq g(U') + \mathrm{trace} \left(\left( \nabla g(U') \right)^T (U-U') \right).$$
Recalling that $\frac{ \partial \mathrm{trace}(X B X^T C)}{\partial X} = C^TXB^T + CXB$, it follows that 
$$\nabla g(U) = -4 \sum_{k=1}^K A_+^k U \Lambda^k.$$
Therefore, 
\begin{align*}
g(U) & \leq g(U') -4 \sum_{k=1}^K \mathrm{trace} \left( \Lambda^k U'^T A_+^k (U-U') \right) \\
& = g(U') -4 \sum_{k=1}^K \mathrm{trace} \left( \Lambda^k U'^T A_+^k U \right) + 4 \sum_{k=1}^K \mathrm{trace} \left( \Lambda^k U'^T A_+^k U' \right)\\
& = -g(U') -4 \sum_{k=1}^K \mathrm{trace} \left( \Lambda^k U'^T A_+^k U \right).
\end{align*}
Therefore, we have shown that $h(U|U')=-g(U') -4 \sum_{k=1}^K \mathrm{trace} \left( \Lambda^k U'^T A_+^k U \right)$ majorizes $g(U)$. 
\end{proof} 

\subsection*{Proof of Proposition~\ref{prop:updateU}.}
\begin{proof}
We can see by inspection that in order to minimize
 \begin{equation}
  \sum_{k=1}^K \| A_+^k -  U  \Lambda^k U^T  \|_F^2
  \label{eq:solveU}
  \end{equation} with respect to an orthogonal matrix $U$,
   it suffices to minimize $g(U)$ in Proposition~\ref{prop:majorize}. Recall from that proposition that $g(U) \leq h(U \mid U') = -g(U') -4 \sum_{k=1}^K \mathrm{trace} \left( \Lambda^k U'^T A_+^k U \right)$.  
Taking a majorization-minimization approach \citep{hunter2004tutorial}, we can decrease the objective of \eqref{eq:solveU} evaluated at $U'$ by solving the optimization problem 
$$\minimize_{U \in \mathbb{R}^{n \times d}, \; U^T U=I}  \left\{ -g(U') -4 \sum_{k=1}^K \mathrm{trace} \left( \Lambda^k U'^T A_+^k U \right) \right\},$$
where $U'$ is the value of $U$ obtained in the previous iteration.  This is equivalent to solving
$$\minimize_{U \in \mathbb{R}^{n \times d}, \; U^T U=I}  \left\{ - \mathrm{trace} \left( \left(  \sum_{k=1}^K \Lambda^k U'^T A_+^k  \right) U \right) \right\},$$
or equivalently, to solving
$$\minimize_{U \in \mathbb{R}^{n \times d}, \; U^T U=I}  \left\{  \left\|   \sum_{k=1}^K A_+^k U' \Lambda^k  - U \right\|_F^2 \right\}.$$
This is an orthogonal Procrustes problem \citep{schonemann1966generalized}, for which the solution is $B C^T$, where the columns of $B$ and $C$ are the left and right singular vectors, respectively, of the matrix   $\sum_{k=1}^K   A_+^k U' \Lambda^k$. 
\end{proof}

\subsection*{Proof of Proposition~\ref{prop:updateLambda}}

\begin{proof}
We see by inspection that in order to solve \eqref{eq:optLambda}, it suffices to solve
 $$\minimize_{\Lambda^1,\ldots,\Lambda^K \in \Delta_+^d} \left\{ \sum_{k=1}^K \left(-2 \mathrm{trace}\left ( U^T A_+^k U \Lambda^k  \right) + \sum_{j=1}^n \left(\Lambda^k_{jj}\right)^2\right) \right\},$$
or equivalently, to solve 
$$ \minimize_{\Lambda_{jj}^k \geq 0  } \left\{   -2  Z^k_{jj} \Lambda^k_{jj}   + \left(\Lambda^k_{jj}\right)^2  \right\}$$
for $j =1,\ldots,n$ and $k=1,\ldots,K$, for $Z^k = U^T A_+^k U$. 
The result follows directly. 
\end{proof}



%
%
%
%
%
%



\subsection*{Proof of Proposition~\ref{prop:null}}

\begin{proof}
Because $U$ is orthogonal and $\Lambda$ is diagonal, we have that %
\begin{align}\label{eqn:note}
 \|A_+^k - U\Lambda U^T\|_F^2 =\|A_+^k\|_F^2 -2\text{trace}(U\Lambda U^TA_+^k) + \sum_{j=1}^d \Lambda_{jj}^2,
\end{align}
where $\Lambda_{jj}$ is the $j$th diagonal element of $\Lambda$. Thus, the optimization problem in  \eqref{eqn:nullproblem+} can be re-written as 
\begin{equation}
\minimize_{{U\in \mathbb{R}^{n\times d},\; U^TU=I , \; \Lambda  \in \Delta_+^d}} \left\{ \left\| {\frac{1}{K} \sum_{k=1}^K A_+^k  -U\Lambda U^T} \right\|_F^2\right\}. 
\end{equation}
Notice that $\frac{1}{K} \sum_{k=1}^K A_+^k$ is the sum of positive semidefinite matrices, and is therefore positive semidefinite. 
The result follows directly from the fact that the truncated singular value decomposition yields the best approximation to a matrix in terms of Frobenius error \citep{eckart1936approximation,johnson1963theorem}.
\end{proof}

\bibliographystyle{Chicago}                           
\bibliography{references.bib} 

\end{document}